\documentclass[aps,onecolumn,floatfix,nofootinbib]{revtex4}
\usepackage{graphicx,epsfig,wrapfig,color}
\usepackage{amsmath}
\usepackage{amssymb}
\usepackage{bm}
\usepackage{dsfont}
\usepackage{wasysym}
\usepackage{color}

\newcommand{\be}{\begin{equation}}
\newcommand{\ee}{\end{equation}}
\newcommand{\ba}{\begin{eqnarray}}
\newcommand{\ea}{\end{eqnarray}}
\newcommand{\sub}[1]{	\begin{subequations}
			#1
		     	\end{subequations} }
\newcommand{\subba}[1]{	\begin{subequations}
			\ba 
			#1 
			\ea
		     	\end{subequations} }
\newcommand{\di}{\!{\rm d}}
\newcommand{\la}{\langle}
\newcommand{\ra}{\rangle}

\begin{document}
\newcommand*{\UConn}{
   Department of Physics, University of Connecticut,
   Storrs, CT 06269-3046, U.S.A.}\affiliation{\UConn}
\title{\boldmath
	$D$-term and structure of point-like and composed spin-0 particles}
\author{Jonathan Hudson}\affiliation{\UConn}
\author{Peter Schweitzer}\affiliation{\UConn}
\date{September 2017}
\begin{abstract}
This work deals with form factors of the energy-momentum tensor (EMT) 
of spin-0 particles and the unknown particle property $D$-term related
to the EMT, and is divided into three parts.\\
\indent
The first part explores free, weakly and strongly interacting theories
to study EMT form factors  with the following findings.
(i) The free Klein-Gordon theory predicts for the $D$-term $D = -1$.
(ii)~Even infinitesimally small interactions can drastically impact $D$.
(iii)~In strongly interacting theories one can encounter large negative $D$ 
	though notable exceptions exist, which includes Goldstone bosons 
	of chiral symmetry breaking.
(iv) Contrary to common belief one cannot arbitrarily add 
	``total derivatives'' to the EMT. Rather the EMT must be defined 
	in an unambiguous way.\\
\indent
The second part deals with the interpretation of the information content
of EMT form factors in terms of 3D-densities with following results. 
(i) The 3D-density formalism is internally consistent.
(ii) The description is subject to relativistic corrections but 
      	those are acceptably small in phenomenologically relevant 
	situations including nucleon and nuclei.
(iii) The free field result $D=-1$ persists when a spin-0
	boson is not point-like but ``heuristically given 
	some internal structure.''\\
\indent 
The third part investigates the question, whether such 
``giving of an extended structure'' can be implemented dynamically, 
and has the following insights. 
(i) We construct a consistent microscopic theory which, in a certain 
	parametric limit, interpolates between extended and point-like 
	solutions.
(ii) This theory is exactly solvable which is rare in $3+1$ dimensions, 
	admits non-topological solitons of $Q$-ball-type, and has a 
	Gaussian field amplitude. 
(iii) The interaction of this theory belongs to a class of logarithmic 
	potentials which were discussed in literature, albeit in different 
	contexts including beyond standard model phenomenology, cosmology, 
	and Higgs physics.
\end{abstract}
%
%
\maketitle




\section{Introduction}
\label{Sec-1:introduction}

The energy momentum tensor operator (EMT) is at the heart of the field 
theoretical description of particles. Through it matter and gauge fields 
couple to gravity, and its matrix elements define fundamental properties 
like mass, spin and the experimentally unknown $D$-term. The latter,
despite being among the most fundamental particle properties 
and although its presence was established in the 1960s when Pagels 
introduced EMT form factors \cite{Pagels}, has received little attention 
for a long time as no practical process was known how to measure EMT 
form factors.

The situation changed in 1990s with the advent of generalized parton 
distribution functions (GPDs) accessible in hard-exclusive reactions 
\cite{Muller:1998fv,Ji:1996ek,Radyushkin:1996nd,Collins:1996fb}.
The second Mellin moments of unpolarized GPDs are related to EMT form 
factors, allowing us to access information about the spin decomposition 
of the nucleon \cite{Ji:1996ek}, the $D$-term \cite{Polyakov:1999gs},
and mechanical properties \cite{Polyakov:2002yz}. The relation of the 
$D$-term to GPDs was further clarified in \cite{Teryaev:2001qm}. The 
potential of GPD studies as a rich source of new information about nucleon 
structure  goes much further \cite{Goeke:2001tz,Diehl:2003ny,
Belitsky:2005qn,Guidal:2013rya,Vanderhaeghen:1998uc}. 

Similarly to electric form factors providing information on the 
electric charge distribution \cite{Sachs}, the EMT form factors offer
insights on the spatial energy density, orbital angular momentum density, 
and the stress tensor \cite{Polyakov:2002yz}. 
The EMT densities not only provide a unique way to gain insights on the 
particle stability and mechanical properties, but also have important
practical applications \cite{Eides:2015dtr}.
For a recent review on the $D$-term we refer to \cite{Hudson:2016gnq}.

The purpose of this work is to provide a comprehensive discussion 
of the EMT and the $D$-term in spin-0 systems.
The goal, besides establishing a benchmark for further studies,
is to focus on clarifying what the $D$-term is and means, undistracted 
by technical details associated with non-zero spin which will be 
addressed elsewhere \cite{Hudson-PS-II}.

The first part of our work is devoted to EMT form factors and the $D$-term.
We explore free, weakly and strongly interacting theories. We first study the 
free field theory case which yields $D=-1$ and provides a point of reference. 
We then discuss how interactions can affect the $D$-term.
We explore the $\Phi^4$ theory as an example of a weakly interacting case
and  show that interactions, even if infinitesimally small, have drastic 
impact on the $D$-term. Hereby we show that in general it is not 
permissible to add total derivatives to the EMT, contrary to common belief. 
Rather such ``improvement terms'' to the EMT operator, if they are needed,
must be chosen with care and require a unique, unambiguous definition. 
In strongly interacting theories, where we consider Goldstone bosons 
of chiral symmetry breaking and nuclei in QCD and $Q$-balls as examples, 
we show that the $D$-terms can have large magnitudes but are always 
negative. The Goldstone bosons are a notable exception in this context:
chiral symmetry dictates $D=-1$ modulo chiral corrections which are 
modest for pions, and somewhat larger for kaons and $\eta$-mesons.
This part contains original results and has partly also review character. 
This is intentional not only to make this work self-contained and place 
our insights in a wider context.
It is also necessary as relevant results from earlier literature were 
rarely (or not at all) discussed in the context of the $D$-term in 
more recent works.

The second part of our work is focused on the interpretation of EMT 
form factors as 3D densities and presents throughout original results.
We first introduce the 3D-density formalism for the spin-0 case
following the work on spin-$\frac12$ systems \cite{Polyakov:2002yz},
demonstrate the consistency, and discuss the limitations of the approach. 
Starting from the notion of a point-like particle, we investigate 
how the EMT properties are affected when the point-like particle
``is given some internal structure'' and ``acquires a finite size.''
These concepts put us in the position to quantify the 
``relativistic corrections'' associated with 3D-densities.
The presence of these corrections is well-known, but the way we quantify 
them is novel and we find them acceptably small for phenomenologically 
relevant cases including nucleon and nuclei (though derived in spin-0 
case, these findings are valid for any spin).
When ``giving'' a particle ``some internal structure'' we initially
proceed heuristically with the remarkable result that the free
field theory result $D=-1$ is preserved when the particle ``acquires'' 
a finite size. We demonstrate that this heuristic picture is fully 
consistent with EMT conservation and other general principles.

In the third part, we address the question whether it is possible to construct 
a microscopic theory where such an internal structure arises from 
dynamics with $D=-1$ and the EMT densities corresponding to what one 
would heuristically expect for a ``smeared out'' point-like particle. 
We show that a Lagrangian can be constructed with an interaction 
known from different contexts in literature. We demonstrate that this
theory describes stable non-topological solitons of $Q$-ball type, 
and show that it can be solved analytically. This by itself is a 
remarkable result, as it is rare to find analytically solvable 
theories in $3+1$ dimensions.

The outline 
	of this work is as follows. The first part in
Sec.~\ref{Sec-2:FFs-of-EMT}
	is focused on EMT form factors, which we define in
	Sec.~\ref{Sec-2a:general-definitions},
	and evaluate in Klein-Gordon theory in
	Sec.~\ref{Sec-2b:free-case}.
	We discuss the weakly interacting case in
	Sec.~\ref{Sec-2c:weak-interaction-case-phi^4},
	consider strongly interacting theories in
	Secs.~\ref{Sec-2d:strong-interaction-case-QCD} 
	and \ref{Sec-2e:strong-interaction-case-Qball},
	and briefly review also higher 
	spin systems in Sec.~\ref{Sec-2f:other-particles}. 
	The second part in
Sec.~\ref{Sec-3:densities}
	deals with the EMT densities. 	
	We introduce the formalism in
	Sec.~\ref{Sec-3a:densities}, 
	compute the EMT densities of a point-like particle in
	Sec.~\ref{Sec-3b:densities-point-like-particle},
	and discuss limitations of the approach in
	Sec.~\ref{Sec-3c:switching-off}.
	We show that the property $D=-1$ persists when a 
	point-like particle is heuristically given an internal 
	structure in Sec.~\ref{Sec-3e:extended-particle}.
	The third part in
Sec.~\ref{Sec-4:Q-ball-log}
	is devoted to the study of a dynamical theory which describes
	a particle whose internal structure corresponds naturally to
	the notion of a smeared-out point-particle with $D=-1$.
	After a brief review of the EMT of $Q$-balls in
	Sec.~\ref{Sec-4a:Qballs-review} which provides the setting,
	we construct and solve the theory in
	Sec.~\ref{Sec-4b:Qball-log-potential},
	before addressing important technical aspects of this theory in
	Sec.~\ref{Sec-4c:boundary-cond-for-log-Qball-theory},
	and indicating potential applications in
	Sec.~\ref{Sec-4d:Qball-log-applications}.
	In 
Sec.~\ref{Sec-4:conclusions} 
	we present our conclusions.
	The Appendices contain remarks on notation and technical details.

\section{\boldmath EMT form factors of spin-0 particles}
\label{Sec-2:FFs-of-EMT}

In this section we define the EMT form factors of a spin-0 particle,
and calculate the EMT form factors and the $D$-term of an elementary 
free spin-0 boson as described by the free Klein-Gordon theory.  
We then discuss what happens to the $D$-term when interactions are 
present and consider both the weak- as well as strong-coupling regime.

\subsection{Formalism and definitions}
\label{Sec-2a:general-definitions}

For a spin-0 particle with mass $m$ the EMT matrix elements are 
described in terms of two form factors \cite{Pagels},
\be\label{Eq:def-EMT-FF}
	\la\vec{p}^{\,\prime\,}|\hat{T}^{\mu\nu}(0)|\vec{p}\,\ra = 
	\frac{P^\mu P^\nu\!}{2}\, A(t) + 
	\frac{\Delta^\mu\Delta^\nu - g^{\mu\nu}\Delta^2}{2}\,D(t) \, ,
\ee
where $\hat{T}^{\mu\nu}(0)$ denotes the EMT operator at space-time 
position zero. The kinematic variables are defined as
\be\label{Eq:variables}
	P^\mu=p^{\mu \,\prime}+p^\mu\, , \;\;\;
	\Delta^\mu=p^{\mu\,\prime}-p^\mu\, , \;\;\; t = \Delta^2 \, .
\ee
The convention for the covariant normalization of one-particle states is 
\be\label{Eq:norm-states}
	\la\vec{p}^{\,\prime}|\vec{p}\,\ra = 2\,E\,(2\pi)^3\,
	\delta^{(3)}(\vec{p}-\vec{p}^{\,\prime}) \, , \;\;\;
	E=\sqrt{\vec{p}^{\,\,2}+m^2}\,.
\ee
Performing the analytic continuation of the form factors
to zero-momentum transfer yields 
\subba{
	\label{Eq:constraint-A(0)}\lim\limits_{t\to 0} A(t) &=& A(0)=1\, ,\\
	\label{Eq:define-D}	  \lim\limits_{t\to 0} D(t) &=& D(0)\equiv D \,.
}
The constraint (\ref{Eq:constraint-A(0)}) is explained by recalling that 
for $\vec{p}\,\to 0$ and $\vec{p}^{\,\prime}\to 0$ only the $00$-component 
remains in Eq.~(\ref{Eq:def-EMT-FF}), and 
$H=\int\di^3x\;\hat{T}_{00}(x)$ is the Hamiltonian of the system 
with $H\,|\vec{p}\,\ra = m\,|\vec{p}\,\ra$ for $\vec{p}\to 0$.
With the conventions (\ref{Eq:def-EMT-FF},~\ref{Eq:norm-states})
(see Appendix~\ref{App-A:notation} for other notations) 
one obtains the constraint $A(0)=1$ in (\ref{Eq:constraint-A(0)}).
It is important to stress that no such constraint exists for the
form factor $D(t)$ such that the $D$-term $D\equiv D(0)$ must be 
determined from experiment. 

For later convenience, let us disentangle the contributions of the 2 
form factors in Eq.~(\ref{Eq:def-EMT-FF}). For that we contract the 
EMT with the symmetric tensors $g^{\mu\nu}$ and $a^{\mu\nu}$ defined as
\be
	a^{\mu\nu} = \frac{P^\mu P^\nu}{P^2}\,, \;\;\; P^2 = 4m^2-t\,.
\ee
Notice that the only other symmetric tensors available in this case are 
proportional to $(P^\mu\Delta^\nu+P^\nu\Delta^\mu)$ or $\Delta^\mu\Delta^\nu$, 
and both are of no use for our purposes since
$\Delta^\mu\la\vec{p}^{\,\prime\,}|\hat{T}_{\mu\nu}(0)|\vec{p}\,\ra =0$
due to EMT conservation.

With $n=g^{\mu\nu}g_{\mu\nu}=4$ denoting the number of space-time
dimensions we obtain
\subba{
	\label{Eq:project-out-A}
	\biggl[(n-1)\,a^{\mu\nu}-g^{\mu\nu}\biggr]
	\la\vec{p}^{\,\prime\,}|\hat{T}_{\mu\nu}(0)|\vec{p}\,\ra 
	= 
	\frac{n-2}{2}\,P^2 A(t),\;\;\;\;\\
   	\label{Eq:project-out-D}
	\biggl[a^{\mu\nu}-g^{\mu\nu}\biggr]
	\la\vec{p}^{\,\prime\,}|\hat{T}_{\mu\nu}(0)|\vec{p}\,\ra 
	=
	\frac{n-2}{2}\,\Delta^2D(t).\;\;\;\;
}
Specifically for $n=3+1$ space-time dimensions we have
\subba{
	\hspace{-8mm}
	A(t) &=& \frac{1}{P^2}\,
	(3\,a^{\mu\nu}-g^{\mu\nu})\;
	\la\vec{p}^{\,\prime\,}|\hat{T}_{\mu\nu}(0)|\vec{p}\,\ra \, ,\\
	D(t)  &=& 
	\frac{1}{\Delta^2}\,
	(\phantom{3\,}a^{\mu\nu}-g^{\mu\nu})\;
	\la\vec{p}^{\,\prime\,}|\hat{T}_{\mu\nu}(0)|\vec{p}\,\ra \, .
}

\subsection{Free field theory case}
\label{Sec-2b:free-case}

It is instructive to start with the free field case. We consider 
the Lagrangian of a non-interacting real spin-0 field 
\be\label{Eq:Lagrangian-free}
	{\cal L} = \frac12\,(\partial_\mu\Phi)(\partial^\mu\Phi) 
	- V_0(\Phi) 
	\, , \;\;\;
	V_0(\Phi) = \frac12\,m^2\Phi^2
\ee
which describes a free spin-0 boson of mass $m$ 
obeying the Klein-Gordon equation
\be\label{Eq:eom}
	(\square + m^2)\,\Phi(x) = 0\,.
\ee
If under parity transformations the field transforms as 
$\Pi\Phi(x)\Pi^{-1}=\pm\Phi(x)$ then the theory describes scalars 
(for $+$) or pseudoscalars (for $-$).
In theories like (\ref{Eq:Lagrangian-free}) the conserved 
canonical EMT operator is symmetric, and given by
\be\label{Eq:EMT-operator}
	\hat{T}^{\mu\nu}(x)=
	(\partial^\mu\Phi)(\partial^\nu\Phi) - g^{\mu\nu}{\cal L} \;,
\ee
where normal ordering is implied. To evaluate the matrix elements 
of the EMT we recall that the free field solutions to the equation 
of motion (\ref{Eq:eom}) are given by
\be
	\Phi(x) = \int\frac{\di^3k}{2\,\omega_k(2\pi)^3}\;\biggl(
	\hat{a}(\vec{k})\,e^{-ikx}+\hat{a}^\dag(\vec{k})\,e^{ikx}\biggr) 
	\, , \;\;\;
	\omega_k=\sqrt{\vec{k}^{\,2}+m^2}
\ee
with creation and annihilation operators satisfying
$	[\hat{a}(k),\hat{a}^\dag(k^\prime)]=2\,\omega_k\,(2\pi)^3\,
	\delta^{(3)}(\vec{k}-\vec{k}^{\,\prime})$
in canonical equal-time quantization. 
The free one-particle states are defined as
$|\vec{p}_{\rm\, free}\ra = \hat{a}^\dag(\vec{p}\,)\,|0\ra$,
and are normalized covariantly according to Eq.~(\ref{Eq:norm-states})
with the trivial vacuum state normalized as $\la 0|0\ra=1$.
The EMT matrix elements can be readily evaluated
\begin{align}
	\la\vec{p}^{\,\prime}_{\rm\, free}|\hat{T}^{\mu\nu}(x)|\vec{p}_{\rm\, free}\ra  
	= 
	e^{i(p^\prime-p)x}\times\biggl\{		
	p^{\prime\mu} p^\nu+p^{\,\mu}p^{\prime\nu} 	
	- g^{\mu\nu}(p^{\,\prime}\cdot p-m^2)\biggl\}\,.
\end{align}
In the notation of Eq.~(\ref{Eq:variables}) 
one has $p^{\,\prime}\cdot p - m^2 = -\,\frac12\,\Delta^2$ and
$p^{\prime\mu}p^\nu+p^{\,\mu}p^{\prime\,\nu}=\frac12(P^\mu P^\nu-\Delta^\mu\Delta^\nu)$
such that 
\be\label{Eq:KG-EMT}
	\la\vec{p}^{\,\prime}_{\rm\, free}|\hat{T}^{\mu\nu}(x)|\vec{p}_{\rm\, free}\ra  
	= 
	e^{i(p^\prime-p)x}\,\frac12\biggl\{P^\mu P^\nu-\Delta^\mu\Delta^\nu
	+g^{\mu\nu}\Delta^2\biggl\} .
\ee
The trivial dependence on the coordinate $x$ is due to translational invariance
$\hat{T}^{\mu\nu}(x)=\exp(i\hat{P}x)\,\hat{T}^{\mu\nu}(0)\exp(-i\hat{P}x)$
where $\hat{P}^\mu = \int\di^3x\,\hat{T}^{0\mu}$ denotes the momentum 
operator. In most definitions one therefore quotes 
$\hat{T}^{\mu\nu}(0)$ as in Eq.~(\ref{Eq:def-EMT-FF}). 

Comparing the result (\ref{Eq:KG-EMT}) with Eq.~(\ref{Eq:def-EMT-FF})
we see that 
\be\label{Eq:KG-FF}
	A(t) = 1, \;\;\; D(t) = -\,1.
\ee
Several remarks are in order.
First, the form factors are constant functions of $t$ as expected for a 
free point-like particle. Second, the constraint $A(0)=1$ in
(\ref{Eq:constraint-A(0)}) is of course satisfied. 
Third, the free Klein-Gordon theory makes 
the unambiguous prediction $D=-\,1$ and the negative sign is in line with
studies in other theoretical frameworks. Fourth, repeating the calculation 
with a complex Klein-Gordon field reveals that a spin-0 particle and its 
anti-particle have the same $D$-term. 

It seems to have been largely overlooked in more recent literature
that in Ref.~\cite{Pagels} not only the notion of EMT form factors 
was introduced for spin-0 and spin-$\frac12$ hadrons and applications
were discussed.
In addition to that in Ref.~\cite{Pagels} also the form factors of a free 
Klein-Gordon particle were quoted. Our result in Eq.~(\ref{Eq:KG-FF}) 
agrees with Ref.~\cite{Pagels}.

The free Klein-Gordon prediction for the $D$-term of a spin-0 particle
sets a reference point for further studies. It is instructive
to examine what happens if one switches on interactions
or the particle is not point-like but extended. We will
investigate these topics in the following.

\subsection{Weakly interacting case}
\label{Sec-2c:weak-interaction-case-phi^4} 

Let us introduce in (\ref{Eq:Lagrangian-free}) a generic interaction,
$V(\Phi) = \frac12\,m^2\Phi^2 + {\cal O}(\lambda)$, characterized by a 
small coupling constant $\lambda\ll 1$ such that it is justified to use 
perturbative methods to solve the theory. In such a situation,
one could naively think the $D$-term would be 
$D_{\rm interacting\,\,naive} =-\,1+{\cal O}(\lambda)$ and reduce to the 
free theory value (\ref{Eq:KG-FF}) for $\lambda\to 0$.
However, this is not the case for two reasons.
(i) 
As a conserved current, the EMT is a renormalization scale invariant operator
so its matrix elements cannot depend on the renormalization scale $\mu$. 
But $\lambda$ acquires in an interacting quantum field theory a dependence
on $\mu$ governed by the $\beta$-function of the theory. Therefore the 
$D$-term must not receive an ${\cal O}(\lambda)$-contribution in a 
perturbative treatment of an interacting theory.  
(ii) 
As no ${\cal O}(\lambda)$-contribution is allowed,
one could then naively think that $D_{\rm interacting\,\,naive} =-\,1$.
However, in general also this is not the case.  We illustrate this point 
considering a specific interacting scalar theory, the $\Phi^4$ theory.

The EMT of the $\Phi^4$ theory was studied in detail in 
Ref.~\cite{Callan:1970ze}. In our context it is instructive 
to review here the findings from Ref.~\cite{Callan:1970ze}, 
see also the works \cite{Coleman:1970je,Freedman:1974gs,Freedman:1974ze,
Lowenstein:1971vf,Schroer:1971ud,Collins:1976vm}.
The theory is defined by
\be\label{Eq:Lagrangian-Phi^4-II}
	{\cal L} = \frac12\,(\partial_\mu\Phi)(\partial^\mu\Phi) 
	- V(\Phi) \, , \;\;\;
	V(\Phi) = \frac12\,m^2\Phi^2 + \frac{\lambda}{4!}\,\Phi^4\,.
\ee
According to the general understanding one can add to the EMT operator 
(\ref{Eq:EMT-operator}) ``any quantity whose divergence is zero and which 
does not contribute to the Ward identities'' \cite{Collins:1976vm}.
(Below we shall see that this general statement has to be formulated 
more carefully.) 
Among possible choices the following ``improvement term'' 
plays a special role  \cite{Callan:1970ze},
\be\label{Eq:EMT-improved-II}
	T^{\mu\nu}_{\rm improve} = 
	T^{\mu\nu}_{\mbox{\footnotesize Eq.(\ref{Eq:EMT-operator})}} +
	\theta^{\mu\nu}_{\rm improve} ,
	\;\;\;\;\;
	\theta^{\mu\nu}_{\rm improve} = -h
	(\partial^\mu\partial^\nu-g^{\mu\nu}\square)\,\phi(x)^2  \,,
	\;\;\;\;\;
	h = \frac14\biggl(\frac{n-2}{n-1}\biggr),
\ee
where $n$ denotes the number of space-time dimensions.
To motivate the improvement term (\ref{Eq:EMT-improved-II}) 
we recall that the coupling of spin-0 fields like 
(\ref{Eq:Lagrangian-free},~\ref{Eq:Lagrangian-Phi^4-II}) 
to gravity is given by an effective action
\be\label{Eq:S-grav-II}
	S_{\rm grav} = \int\di^4x\;\sqrt{-g}\biggl(
	\frac12\,g^{\mu\nu}(\partial_\mu\Phi)(\partial_\nu\Phi)- V(\Phi)
	-\frac12\,h\,R\,\Phi^2\biggr)
\ee
where $-\frac12\,h\,R\,\Phi^2$ is a non-minimal coupling term, $R$ is 
the Riemann scalar, $g$ denotes the determinant of the metric,
and it is understood that gravity is treated to lowest order. 
From (\ref{Eq:S-grav-II}) one obtains the EMT operator via
\be\label{Eq:EMT-from-gravity-II}
	T_{\mu\nu} = \frac{2}{\sqrt{-g}}\,
	\frac{\delta S_{\rm grav}}{\delta g^{\mu\nu}}\,.
\ee
Omitting the non-minimal term in (\ref{Eq:S-grav-II}) yields a correct 
description of a scalar field theory (minimally) coupled to a gravitational 
background field, and one recovers from (\ref{Eq:EMT-from-gravity-II}) the 
canonical EMT operator (\ref{Eq:EMT-operator}). Keeping the non-minimal 
term yields the improved EMT (\ref{Eq:EMT-improved-II}).
(In flat space the Riemann scalar $R$ vanishes, but its variation 
with respect to the metric is nevertheless non-zero.)

In classical theory, the improvement term with the particular value for 
$h$ in (\ref{Eq:EMT-improved-II}) is fixed by requiring the kinetic energy 
in (\ref{Eq:S-grav-II}) to be conformally invariant: with this improvement
term the trace ${T^{\mu}}_\mu = m^2\Phi(x)^2$ which preserves conformal symmetry 
of the classical theory in the limit where $m$ vanishes. On quantum level, 
the conformal symmetry is broken, but the improvement term is required to 
make Greens functions of the renormalized fields with an insertion of the 
improved EMT (\ref{Eq:EMT-improved-II}) finite. More precisely, the value 
for $h$ in  (\ref{Eq:EMT-improved-II}) removes UV divergences up to 
three-loops in dimensional regularization.
The four-loop expression for $h$ would acquire in addition to the result
quoted in (\ref{Eq:EMT-improved-II}) a contribution proportional to $(n-4)^3$ 
needed to cancel pole contributions in dimensional regularization. However, 
the overall shape of the improvement term and the independence of $h$ on 
the renormalized coupling, mass, renormalization scale $\mu$ remain to
all orders \cite{Collins:1976vm}. 

To compute the $D$-term in $\Phi^4$ theory it is therefore sufficient to
investigate the effect of the improvement term at tree-level:
loop corrections produce UV divergences which the 
improvement term (\ref{Eq:EMT-improved-II}) removes
\cite{Callan:1970ze,Coleman:1970je,Freedman:1974gs,Freedman:1974ze,
Lowenstein:1971vf,Schroer:1971ud,Collins:1976vm}, and due to the
renormalization scale invariance of the EMT operator the final
result must not be altered by ${\cal O}(\lambda)$-corrections. 
Evaluating the improvement operator at tree-level yields
\be\label{Eq:improvement-term-evaluate}
	\la\vec{p}^{\,\prime}_{\rm\, free}|\,\hat{\theta}^{\mu\nu}_{\rm improve}(x)\,
	|\vec{p}_{\rm\, free}\ra  
	= 2\,h\,e^{i(p^\prime-p)x}\,
	\biggl\{\Delta^\mu\Delta^\nu-g^{\mu\nu}\Delta^2\biggl\} \, .
\ee

There is no effect on $A(t)$. This is expected because $A(0)=1$ is fixed 
from general principles and one obtains this result already without
including any improvement term, see Sec.~\ref{Sec-2b:free-case}. The
inclusion of the improvement term therefore must not, and does not, 
spoil the general constraint (\ref{Eq:constraint-A(0)}).

The situation is different for the $D$-term which interestingly is
affected. From Eq.~(\ref{Eq:improvement-term-evaluate}) we obtain 
\be\label{Eq:D-interactong-with-improvement-term-II}
	D_{\rm interacting\; improved} = -1 + 4\,h \,.
\ee
With $h=\frac16$ in $n=3+1$ space-time dimensions we obtain
$D_{\rm interacting\; improved} = -\,\frac13$.
This a remarkable result. Even infinitesimally weak interactions
can have a drastic effect on the value of the $D$-term.
This insightful observation deserves several comments.

First, 	
adding total derivatives to the EMT leaves
$P^\mu\equiv \int\di^3x T^{0\mu}$ and other Poincar\'e group generators
unaffected, i.e.\ it does not impact the particle mass or spin. But 
we see that $D$ in general is sensitive to adding total derivatives:
the improvement term is one such total derivative.
The $D$-term is a measurable quantity, even though challenging 
to infer from experiment.
This means in general one 
cannot add total derivatives to the EMT at will, 
contrary to common belief. When this happens to be necessary
(Belifante procedure in Dirac case, $\Phi^4$ theory)
it is crucial to establish a unique definition for improvement
term(s) as dictated by the general properties of the theory,
in order to ensure a uniquely defined $D$-term.

Second, 
when dealing with a free massive field theory case, there is no 
criterion to motivate and uniquely define a specific improvement 
term. In lack of such a criterion we conclude that in free scalar 
theory $D=-1$, Eq.~(\ref{Eq:KG-FF}). 
This is an unambiguous prediction of the 
free Klein-Gordon theory (minimally coupled to gravity), analog 
to the anomalous magnetic moment $g=2$ predicted from free Dirac 
theory (minimally coupled to an electromagnetic background field).

Third, in $\Phi^4$ theory we deal with an interacting quantum 
field theory  which has to be renormalized. In this case 
the unique improvement term (\ref{Eq:EMT-improved-II}) ensures that Greens 
functions with an insertion of the improved EMT are finite. This guarantees 
the ``renormalizability of the combined theory of gravity and matter, 
with gravity treated to lowest order and the self-interactions of 
matter [in $\Phi^4$ theory] to all orders'' 
\cite{Collins:1976vm}.
The inclusion of the improvement term has a drastic effect on the
$D$-term. Assuming even an infinitesimally small coupling constant 
$\lambda\lll 1$ (such that calculations to three or fewer loops are
sufficient) we have $D_{\rm interacting\;improved} =-\,\frac13$ instead of the 
value $-1$ in the free theory.\footnote{\label{footnote-massless-case}%
	For completeness we remark that in the conformally invariant
	massless free scalar theory, one also has to introduce the 
	improvement term (\ref{Eq:EMT-improved-II}) to restore 
	${T^\mu}_{\!\mu}=m^2\Phi(x)^2\to0$ and recover a divergenceless 
	(conserved) conformal current. Thus, in the massless free case
	we also have $D=-\,\frac13$. 
	At this point one may wonder whether the improvement term 
	(\ref{Eq:EMT-improved-II}) should also be added 
	in the massive free Klein-Gordon theory. Then the $D$-term
	would exhibit a smooth behavior when $m$ goes to zero. This would
	certainly be a legitimate step, though there is in general no 
	reason to expect necessarily a smooth behavior of particle
	properties in a limit such as $m\to0$. However, one may also
        invoke arguments which support that $D=-1$ is a consistent 
	result in the massive free case, see 
	App.~\ref{App-B:point-like-particle-consistency}.
	At the end we shall briefly review which definition of the EMT
	is appropriate in App.~\ref{App-C:canon-vs-conform}.}
This clearly demonstrates that the $D$-term is highly sensitive to 
interactions and the dynamics.

Fourth, the renormalizability of the $\Phi^4$ theory has been studied
in weak curved gravitational background fields, and the same improvement 
term (\ref{Eq:EMT-improved-II}) is required \cite{Brown:1980qq}, which 
means $D=-\,\frac13$ in weakly interacting $\Phi^4$ theory in 
presence of gravity. As no quantum theory of gravity is known, it is of 
course also not known whether (\ref{Eq:EMT-improved-II}) would ensure 
renormalizability if quantum gravity effects were included. 
At this point one might be tempted to think that gravity is far too
weak to be of relevance in particle physics. However, the lesson we
learned is that even infinitesimally small interactions in
$\Phi^4$ theory can impact the $D$-term. So why not infinitesimally 
small gravitational~interactions?

Fifth, the $D$-term emerges to be strongly sensitive to interactions. One 
must consistently include all forces, perhaps even gravity, to determine 
the true improvement term and the ``true'' value of the $D$-term.
These issues are beyond the scope of our work as is the very question whether 
a non-trivial $\Phi^4$ theory actually exists \cite{Callaway:1988ya}. 

The above arguments certainly do not apply to theories which have to be 
solved in non-perturbative regime. At this point one may therefore wonder 
how the $D$-terms of spin-0 particles are affected in strongly 
interacting theories. We shall discuss two examples in 
the next sections, QCD and $Q$-balls.

\subsection{Strongly interacting theory, QCD}
\label{Sec-2d:strong-interaction-case-QCD}

It is not possible to tell what the $D$-term would be in a strongly
interacting $\Phi^4$-theory, where the perturbative expansion indicated
in Eq.~(\ref{Eq:D-interactong-with-improvement-term-II}) would be 
inappropriate. Fortunately, the $D$-terms can be computed for a special
class of spin-0 particles in a much more relevant and realistic
strongly interacting theory, QCD. This is possible for pions, kaons
and $\eta$-meson, the Goldstone bosons of chiral symmetry breaking
by exploring low energy theorems. The results were already obtained in
1980, but have not been discussed in the context of the physics of the
$D$-term. It is therefore of interest to review them here.

In Refs.~\cite{Novikov:1980fa,Voloshin:1980zf} the charmonium decays 
$\psi^\prime \to J/\psi\,\pi\,\pi$ were studied. 
The description of these decays requires the matrix elements 
$\la\pi^{\,\prime}\pi\,|\hat{T}^{\mu\nu}(0)|0\,\ra$, 
or $\la\pi^{\,\prime\,}|\hat{T}^{\mu\nu}(0)|\pi\,\ra$ after applying
crossing symmetry. Similar matrix elements enter also the description
of a hypothetical light Higgs boson decay \cite{Ellis:1975ap} into two 
pions which was discussed at some point in the past in literature 
\cite{Donoghue:1990xh}.

Chiral symmetry uniquely determines the interactions of soft pions.
In Refs.~\cite{Novikov:1980fa,Voloshin:1980zf} the following low energy 
theorem was derived which, in our notation, is given by 
\be\label{Eq:low-energy-theorem}
	\la\pi(\vec{p}^{\,\prime\,})|\hat{T}^{\mu\nu}(0)|\pi(\vec{p}\,)\,\ra 
	= \frac12\biggl(P^\mu P^\nu-\Delta^\mu\Delta^\nu
	+g^{\mu\nu}\Delta^2\biggr) + {\cal O}(E^4)\, .
\ee
Here $E$ is the soft scale associated with the soft momenta of the Goldstone 
bosons or their masses, i.e.\ generically $E \sim {\cal O}(p,p^\prime,m_\pi)$.
From (\ref{Eq:low-energy-theorem}) we read off (notice the first term on the 
right-hand side of (\ref{Eq:low-energy-theorem}) is already $E^2$) 
\be\label{Eq:D-term-soft-pion}
	D_h = -1 + {\cal O}(E^2) \, ,\;\;\; h = \pi,\;K,\;\eta,
\ee
where we added that the same result is obtained also for kaons and the
$\eta$-meson. This is a remarkable result. In the soft pion limit chiral 
symmetry dictates that the form factors of the EMT and the $D$-term of the 
light octet mesons coincide (at small values of $-t\sim m_\pi^2 \sim E^2$) 
with the free-field case in Eq.~(\ref{Eq:KG-FF}), despite the fact that 
we deal with strongly interacting particles. 
Notice, however that the Goldstone bosons have 
no internal structure to the considered order in the soft scale in 
Eqs.~(\ref{Eq:low-energy-theorem},~\ref{Eq:D-term-soft-pion}), which 
makes it plausible why the free field value (\ref{Eq:KG-FF}) is naturally 
recovered.
In particular, this implies that
\be\label{Eq:D-term-pion-chiral-limit}
	\lim_{E\;\to\; 0} \;D_h = -1 \, ,\;\;\; h = \pi,\;K,\;\eta.
\ee
This result was derived independently from a soft-pion theorem for
pion GPDs in Ref.~\cite{Polyakov:1999gs}. 
At this point one may wonder why no improvement term analog 
to (\ref{Eq:EMT-improved-II}) was added, which would be relevant 
in massless case, see footnote~\ref{footnote-massless-case}. 
However, the answer is that such an improvement term is forbidden
as it violates chiral symmetry \cite{Voloshin:1982eb,Leutwyler:1989tn}.

The chiral properties of the EMT form factors $A_i(t)$ and
$D_i(t)$ for $i=\pi,\;K,\;\eta$ were studied beyond the chiral limit
and evaluated in chiral perturbation theory to order ${\cal O}(E^4)$ 
in Ref.~\cite{Donoghue:1991qv}. We quote here only the results for
the $D$-terms \cite{Donoghue:1991qv} which are given by
\begin{subequations}
\ba
	D_\pi &=& - 1 
		+ 16\,a\,\frac{m_\pi^2}{F^2}
		+    \frac{m_\pi^2}{F^2}\,I_\pi
		-     \frac{m_\pi^2}{3F^2}\,I_\eta + {\cal O}(E^4)
		\label{Eq:chiral-prediction-RAW-pi}\\
	D_K \! &=& - 1 
		+ 16\,a\,\frac{m_K^2}{F^2}
		+     \frac{2m_K^2}{3F^2}\,I_\eta + {\cal O}(E^4)
		\label{Eq:chiral-prediction-RAW-K}\\
	D_\eta &=& - 1 
		+ 16\,a\,\frac{m_\eta^2}{F^2}
		-     \frac{m_\pi^2}{F^2}\,I_\pi
		+     \frac{8m_K^2}{3F^2}\,I_K 
		+     \frac{4m_\eta^2-m_\pi^2}{3F^2}\,I_\eta + {\cal O}(E^4)
		\label{Eq:chiral-prediction-RAW-eta}
\ea
where
\be\label{Eq:details}
	a = L_{11}(\mu)-L_{13}(\mu) \, , \;\;\;
	I_i = \frac{1}{48\pi^2}\,\biggl({\rm log}\frac{\mu^2}{m_i^2}-1\biggr)
	\, , \;\;\; i = \pi, \; K, \; \eta \, ,
\ee
\end{subequations}
and $F$ denotes the pion decay constant $F \simeq 93\,{\rm MeV}$. The 
expansion parameter in chiral perturbation theory is often associated with 
the dimensionless ratio $E^2/(4\pi F)^2$ where $(4\pi F)^2\sim 1\,{\rm GeV}^2$.
In Eq.~(\ref{Eq:details}) the renormalization scale $\mu$ appears, which is
arbitrary because changes in $\mu$ are absorbed by appropriate redefinitions
of the low energy constants $L_{11}$ and $L_{13}$. This reflects the fact
that the EMT is a renormalization scale invariant operator.
Notice also that to the order considered in 
(\ref{Eq:chiral-prediction-RAW-pi}--\ref{Eq:details})
which corresponds to ${\cal O}(E^6)$ in Eq.~(\ref{Eq:low-energy-theorem})
the form factors exhibit a $t$-dependence, which signals that the
Goldstone bosons acquire an internal structure.

This allows one to make more realistic predictions for the $D$-terms 
than the chiral limit prediction (\ref{Eq:D-term-pion-chiral-limit}).
The values of the low energy constants were estimated \cite{Donoghue:1991qv}
as $L_{11}(1\,{\rm GeV}) = (1.4\mbox{--}1.6)\times 10^{-3}$ and 
$L_{13}(1\,{\rm GeV}) = (0.9\mbox{--}1.1)\times 10^{-3}$ using 
the meson dominance model (lower values) 
and dispersion relation technics (higher values). 
This yields
\subba{
	D_\pi	&=&  -0.97 \pm 0.01 \,,\label{Eq:chiral-prediction-pi}\\
	D_K \!	&=&  -0.77 \pm 0.15 \,,\label{Eq:chiral-prediction-K}\\
	D_\eta 	&=&  -0.69 \pm 0.19 \,,\label{Eq:chiral-prediction-eta}
}
where the uncertainties are due to 
$\delta L_{11} = \delta L_{13} = 0.2\times 10^{-3}$, the use of the physical 
value of the pion decay constant $F = 93\,{\rm MeV}$ \cite{Donoghue:1991qv} 
vs chiral limit value $F = 88\,{\rm MeV}$ \cite{Kubis:1999db},
and a heuristic estimate of higher order chiral corrections
proportional to $E^4/(4\pi F)^4$ with $E$ the respective meson mass.
All these uncertainties are added in quadrature. 
Chiral interactions alter the soft theorem result $D=-1$, and 
are not unexpectedly more sizable for heavier mesons. 
However, the $D$-terms remain negative.

For completeness we remark that the effects of the electromagnetic 
interaction on the EMT form factors of charged and neutral pions 
were investigated in \cite{Kubis:1999db}.
More recently pion EMT form factors were studied in chiral quark
models, where definite predictions for the low energy constants
can be made \cite{Megias:2004uj}.

The quark contribution to pion EMT form factors was also studied
in lattice QCD for pion masses in the range 
$550\,{\rm MeV} \le m_\pi\le 1090\,{\rm MeV}$ for lattice spacings
0.07--0.12$\,$fm and spatial lattice sizes 1.6--2.2$\,$fm
\cite{Brommel:2005ee,Brommel:2007zz}. 
The quark contribution to the $D$-term was found to be,
see Table~7.3 in  \cite{Brommel:2007zz},
\be
	D^Q_\pi = - (0.264 \pm 0.032)
\ee 
at a renormalization scale of $2\,{\rm GeV}$ in 
$\overline{\mbox{\footnotesize MS}}$ scheme.
The error includes the statistical accuracy of the lattice
simulations combined with an estimate of uncertainties due to 
the extrapolation procedure (to physical pion masses and $t=0$). 
Finite volume effects were noticed but could not be quantified 
as systematic uncertainties \cite{Brommel:2005ee,Brommel:2007zz}.
It is not possible to confront this result with the prediction 
(\ref{Eq:chiral-prediction-pi}) from chiral perturbation theory
because $D^Q_\pi = - (0.264 \pm 0.032)$ is only a partial result 
(currently no information from lattice QCD is available on the 
gluonic contribution to the $D$-term of the pion or any other hadron).
In addition it is difficult to reliably quantify the uncertainty due
to extrapolation from the pion mass region above $550\,{\rm MeV}$ to
the physical point. It will be interesting to see new lattice
calculations on present-day state-of-the-art lattices where physical
pion masses can be handled.

The light pseudoscalar octet mesons are an exception, since they
are Goldstone bosons of chiral symmetry breaking. For other hadrons 
no low energy theorems exist which would allow to predict their $D$-terms, 
and one may in general obtain much different numerical values for $D$.
This is nicely illustrated by studies of nuclei. In general the description 
of nuclei in QCD is rather complex, and certainly no easier than that of 
Goldstone bosons and any other hadron. However, the saturation property and 
short range of the ``residual'' nuclear forces make it possible to predict 
gross features of nuclear $D$-terms.

Both properties are well-captured in the liquid drop model which was 
explored to study nuclear $D$-terms \cite{Polyakov:2002yz}.
Of course only ground states of even-even nuclei 
(even number of protons $Z$ and even number of neutrons $N$) are 
``guaranteed'' to have spin zero. But spin effects play no role in 
the liquid drop model. Interestingly, nuclear radii grow as $A^{1/3}$ 
and nuclear masses as $A$ with the mass number $A=N+Z$. But nuclear 
$D$-terms, due to the surface tension in the liquid drop model,
are negative and show a far stronger dependence $D\propto A^{7/3}$ 
\cite{Polyakov:2002yz}.
Numerical calculations in the Walecka model for selected $J^\pi=0^+$
isotopes ($^{12}$C, $^{16}$O, $^{40}$Ca, $^{90}$Zr, $^{208}$Pb) 
were presented in \cite{Guzey:2005ba}. The $D$-terms were found 
negative. For nuclei heavier then $^{12}$C it was found 
$D \propto A^{2.26}$ in good agreement with \cite{Polyakov:2002yz}. 
For completeness we remark that 
in Ref.~\cite{Liuti:2005qj} a different $A$-behavior was found.

Let us summarize what we know about the $D$-terms of spin-0 hadrons.
For the Goldstone bosons of chiral symmetry breaking in strong 
interactions one can explore low energy theorems and chiral 
perturbation theory to predict that $D = -1$ modulo chiral corrections 
which make the $D$-term less negative, but do not
change its sign. $D$-terms of nuclei are also negative,
much more sizable than those of the light pseudoscalar mesons
and strongly grow with the mass number as $D\propto A^{7/3}$
which can be tested in experiments on hard exclusive reactions 
off nuclei \cite{Polyakov:2002yz}.

\subsection{Strongly interacting theory, $Q$-balls}
\label{Sec-2e:strong-interaction-case-Qball}

Another example of a strongly interacting theory of scalar particles 
is the $Q$-ball system \cite{Coleman:1985ki}, see also 
\cite{Friedberg:1976me,Lee:1991ax}. In this section we briefly review 
the $Q$-ball theory and quote some results regarding the $D$-term
from \cite{Mai:2012yc,Mai:2012cx,Cantara:2015sna}. More details about
$Q$-balls will be provided in Sec.~\ref{Sec-4b:Qball-log-potential} 
where we will explore the $Q$-ball framework for further applications.

$Q$-balls are solitons in scalar theories with a global symmetry 
where a ``suitable potential'' satisfies certain conditions.
The theory can be formulated in terms of one complex scalar field,
or equivalently in terms of two real scalar fields which we shall
choose to do here. The Lagrangian and the equations of motion are
given by
\be\label{Eq:Lagrangian-Qball}
	{\cal L} 
	= \frac12(\partial_\mu\Phi_1)(\partial^\mu\Phi_1) 
	+ \frac12(\partial_\mu\Phi_2)(\partial^\mu\Phi_2) 
	- V \, , \;\;\;
	\square \Phi_i(x) + \frac{\partial V}{\partial\Phi_i}
	= 0\,, \;\;\; i = 1,\,2, 
\ee
with a potential $V$ such that the theory is invariant under global
continuous SO(2) symmetry transformations ($\beta\in\mathds{R}$)
\be\label{Eq:Qball-symmetry}
	\left(\begin{matrix} \Phi_1 \\ \Phi_2 \end{matrix}\right) \to
	\left(\begin{matrix} 	\cos\beta & -\sin\beta \\ 
				\sin\beta &  \cos\beta \end{matrix}\right) 
	\left(\begin{matrix} \Phi_1 \\ \Phi_2 \end{matrix}\right) \, . 
\ee
The global symmetry implies a conserved Noether current 
$J^\mu=\Phi_1\partial^\mu\Phi_2-\Phi_2\partial^\mu\Phi_1$. The associated 
conserved charge $Q=\int\di^3x\,J^0(x)$ is instrumental for the existence 
of the soliton solutions which are, in their rest frames, of the type 
\be\label{Eq:ansatz-Qball}
   \left(\begin{matrix}\Phi_1(\vec{x},t)\\ \Phi_2 (\vec{x},t)\end{matrix}\right)
   = \left(\begin{matrix} \cos(\omega t) \\ \sin(\omega t) \end{matrix}\right) 
	\,\phi(r)\, ,
\ee
where $r=|\vec{x}^{ }|$ 
and $\omega$ is bound by $\omega_{\rm min}^2<\omega^2<\omega_{\rm max}^2$.
The limiting frequencies are defined in terms of the properties of the 
potential $V$, with $V$ understood as a function of the radial field $\phi(r)$,
as follows
\be\label{Eq:condition-for-existence}
	0 < 	\omega_{\rm min}^2 
	\equiv	\min\limits_\phi \biggl[\frac{2\,V(\phi)}{\phi^2} \biggr] 
	  <	\omega_{\rm max}^2 
	  = V^{\prime\prime}(\phi)\biggl|_{\phi=0} \,.
\ee
Notice that $m=\omega_{\rm max}$ defines the mass of the elementary
quanta of the fields $\Phi_1$ and $\Phi_2$. The solutions satisfying 
(not satisfying) the equivalent conditions 
\be\label{Eq:classical-stability}
        \frac{\di}{\di\omega}\left(\frac{M}{Q}\right) \ge 0
         \;\;\; \Leftrightarrow \;\;\;\frac{\di Q}{\di\omega} \le 0
         \;\;\; \Leftrightarrow \;\;\;\frac{\di^2 M}{\di Q^2} \le 0\;,
\ee
are stable (unstable) with respect to small fluctuations 
\cite{Friedberg:1976me,Lee:1991ax}.
The point where the inequalities in (\ref{Eq:classical-stability})
become equalities defines the critical frequency $\omega_c$, i.e.\
for instance $Q^\prime(\omega)=0$ at $\omega=\omega_c$.
The solutions are absolutely stable if $M < m\,Q$ where $m$ denotes 
the mass of the elementary fields \cite{Lee:1991ax}. 

In the $Q$-ball system a general analytical proof was formulated 
that $D<0$ for any suitable potential \cite{Mai:2012yc}. 
It was also shown that the numerical values of the $D$-terms can 
span orders of magnitude. For that the suitable, often studied 
(non-renormalizable, effective) theory was used with the sextic 
potential $V_6 = A\,\phi^2 - B\,\phi^4 + C\,\phi^6$ with 
$\phi^2=\Phi_1^2+\Phi_2^2$ and positive $A,B,C$ satisfying 
$0<\zeta\equiv B^2/(4AC)<1$ \cite{Coleman:1985ki}. For this potential 
$\omega_{\rm min}^2=2A(1-\zeta)$ and $\omega_{\rm max}^2=2A$. For the 
parameters $A=1.1,\;B=2.0,\;C=1.0$ it was found $|D| \ge |D_c|$ 
with $D_c = -113.4$ numerically close to the critical frequency 
$\omega_c=1.38$ \cite{Mai:2012yc}. For $\omega$ not in the 
vicinity of $\omega_c$ the $D$-terms are becoming quickly 
more and more negative.

In the ``$Q$-ball limit'' 
$\varepsilon_{\rm min}\equiv\sqrt{\omega^2-\omega_{\rm min}^2}\to0$ 
one deals with absolutely stable well-localized solitons 
\cite{Coleman:1985ki} characterized by constant charge density 
in their interior, whose sizes grow as $\varepsilon_{\rm min}^{-4}$,
and the masses and charges grow as $\varepsilon_{\rm min}^{-6}$. 
The most spectacular growth, however, is exhibited by the $D$-term 
which behaves as $D\propto \varepsilon_{\rm min}^{-14}$ in this limit
\cite{Mai:2012yc}.

In the opposite ``$Q$-cloud limit'' 
$\varepsilon_{\rm max}\equiv\sqrt{\omega_{\rm max}^2-\omega^2}\to0$ 
\cite{Alford:1987vs} the solutions become infinitely dilute, 
diffuse and disintegrate into a cloud of free $Q$-quanta. 
In this limit mass, charge, and mean radii of the solutions 
diverge as $\varepsilon_{\rm max}^{-1}$. Again, the $D$-term is the
property exhibiting the strongest pattern of divergence with
$D\propto \varepsilon_{\rm max}^{-2}$ \cite{Cantara:2015sna}. 
Interestingly, in the $Q$-cloud limit the sextic term in $V_6$
becomes irrelevant (in the sense of critical phenomena), and after a
suitable rescaling one  deals with a (complex) $\Phi^4$ theory continued
analytically to a negative coupling constant $\lambda$ 
\cite{Cantara:2015sna}.

$Q$-balls can have also excited states (all with spin zero and
positive parity as the ground state) which are unstable and have 
also negative $D$-terms. The solution $\phi(r)$ of the $N^{\rm th}$ 
excitation exhibits $N$ nodes (ground state has no node).
For a fixed frequency $\omega$ the mass and charge of the $N^{\rm th}$ 
excitation scale as $N^3$, while the $D$-term scales as $N^8$ 
\cite{Mai:2012cx}. 

The $Q$-ball system confirms that $D$-terms of spin-0 particles can 
deviate significantly from the free-field theory result $D = -1$ 
though the negative sign of the $D$-term is preserved.
The $Q$-ball results also strongly support the observation that 
the $D$-term is the particle property which is most sensitive 
to the details of the dynamics of a theory.

\subsection{Particles with higher spins}
\label{Sec-2f:other-particles}

We remark that also the $D$-terms of particles with non-zero spin
were investigated in a variety of theoretical frameworks and models. 
In all cases the $D$-terms were found negative, including nucleon 
(spin $\frac12$)
\cite{Ji:1997gm,Petrov:1998kf,Schweitzer:2002nm,Goeke:2007fp,
Goeke:2007fq,Cebulla:2007ei,Kim:2012ts,Pasquini:2014vua,
Hagler:2003jd}, photon (spin 1) \cite{Gabdrakhmanov:2012aa}, and
$\Delta$-resonance (spin $\frac32$) \cite{Perevalova:2016dln}. 
Notice that no analog of the low energy theorem (\ref{Eq:low-energy-theorem}) 
exists for hadrons other than Goldstone bosons. Therefore, 
chiral perturbation theory cannot predict the $D$-term of 
e.g.\ the nucleon, though it can make predictions on the 
small-$t$ dependence of the EMT form factors \cite{Chen:2001pv}.

\section{3D EMT densities}
\label{Sec-3:densities}

In this section we introduce the notion of 3D-densities of the EMT,
apply it to the case of a free point-like particle, and demonstrate its 
consistency. We show that the description is physically well-formulated 
and justified in the heavy mass limit.
We then ``give'' the particle a finite size. Hereby we initially proceed
in a heuristic way. The finite size naturally introduces an additional scale 
in the theory, which is required to formulate adequately the heavy mass limit. 
We show that the property $D=-1$ is then still preserved. 
Finally we demonstrate that it is possible to construct dynamical 
microscopic theories which describe extended particles where the 
free field property $D=-1$ is preserved.

\subsection{Static EMT and definitions}
\label{Sec-3a:densities}

The information content associated with EMT form factors can be 
interpreted in analogy to the electromagnetic form factors \cite{Sachs} 
in the Breit frame which is characterized by $\Delta^0=E^\prime-E=0$. 
In this frame, one defines the static energy-momentum tensor
as \cite{Polyakov:2002yz}
\be\label{Def:static-EMT}
    T_{\mu\nu}(\vec{r}\,) =
    \int\frac{\di^3\Delta}{2E(2\pi)^3}
    \;\exp(i\vec{\Delta}\vec{r}\,)\;
    \la \vec{p}^{\,\prime}|\hat{T}_{\mu\nu}(0)|\vec{p}\,\ra \,,
\ee
where $E=E^\prime=\sqrt{m^2+\vec{\Delta}^2/4}$.
This provides information on the energy density $T_{00}(\vec{r}\,)$ and 
the stress tensor $T_{ik}(\vec{r}\,)$. The $T_{0k}(\vec{r}\,)$ 
components vanish in the spin-0 case.

The energy density yields the particle mass according to
$m=\int\di^3r\,T_{00}(\vec{r}\,)$, which implies the constraint 
(\ref{Eq:constraint-A(0)}).
The stress tensor is described in terms of two 
functions, the distribution of shear forces $s(r)$ and pressure $p(r)$,
\be\label{Eq:T_ij-pressure-and-shear}
    T_{ij}(\vec{r}\,)
    = s(r)\left(e_r^ie_r^j-\frac 13\,\delta_{ij}\right)
        + p(r)\,\delta_{ij}\, , 
\ee
where $\vec{e}_r=\vec{r}/r$ denotes the radial unit vector
and $r=|\vec{r}\,|$. The EMT conservation, $\partial^\mu\hat{T}_{\mu\nu}=0$, 
implies for the static stress tensor $\nabla^i T^{ij}(\vec{r}\,)=0$
from which one can derive two helpful relations. First, $p(r)$ and $s(r)$
are connected by 
\be\label{Eq:diff-eq-s-p}
    \frac23\;\frac{\partial s(r)}{\partial r\;}+
    \frac{2s(r)}{r} + \frac{\partial p(r)}{\partial r\;} = 0\;.
\ee
Second, the pressure $p(r)$ must satisfy the von Laue condition 
\cite{von-Laue,BialynickiBirula:1993ce}, which is a necessary 
(but not sufficient) condition for stability,
\be\label{Eq:stability}
    \int\limits_0^\infty \!\di r\;r^2p(r)=0 \;.
\ee
Owing to Eq.~(\ref{Eq:diff-eq-s-p}) the $D$-term can be expressed in 
two different ways in terms of shear and pressure distributions as
\be\label{Eq:D-from-s(r)-and-p(r)}
         D \stackrel{\displaystyle(a)}{=}-\,\frac{4m}{15}\int\di^3r\;r^2\, s(r)
           \stackrel{\displaystyle(b)}{=}m \int\di^3 r\;r^2\, p(r)\;.
\ee

The concepts of ``mechanical stability'' \cite{Polyakov:2002yz}
impose stability criteria on the densities in the classical theory 
which can be introduced also in quantum field theory and imply for 
the EMT densities \cite{Perevalova:2016dln}
\be\label{Eq:local-requirements-stability}
	(a) \;\; T_{00}(r) \ge 0 \, , \;\;\;\;
	(b) \;\; \frac23\,s(r) + p(r) \ge 0 \,.
%
%
\ee

For practical applications it is helpful to derive the explicit 
expressions for the densities in terms of the form factors and 
demonstrate their consistency. For that we recall that in the 
Breit frame $P^\mu=(P^0,0,0,0)$ and $\Delta^\mu=(0,\vec{\Delta})$. 
With this we obtain from (\ref{Def:static-EMT}) for the energy 
density and the stress tensor the results
\subba{
	\label{Eq:static-EMT-T00}
    	T_{00}(r) &=&
    	m^2\int\frac{\di^3\Delta}{E(2\pi)^3}
    	\;e^{i\vec{\Delta}\vec{r}}\;\biggl[A(t)-\frac{t}{4m^2}(A(t)+D(t))\biggr]\\
	\label{Eq:stress-tensor-1}
    	T_{ij}(\vec{r}\,) &=&
    	\frac{1}{2}\,\int\frac{\di^3\Delta}{2E(2\pi)^3}
    	\;e^{i\vec{\Delta}\vec{r}}\;
    	\biggl[\Delta_i\Delta_j - \delta_{ij} \vec{\Delta}^2 \biggr]\,D(t) \,.
}
From Eq.~(\ref{Eq:stress-tensor-1}) we can project out the 
expressions for the pressure and shear forces, namely
\subba{
	\label{Eq:pressure-1}
    	p(r) &=& 
	\frac{1}{3}\int\frac{\di^3\Delta}{2E(2\pi)^3}
	\;e^{i\vec{\Delta}\vec{r}}\;D(t)\,\biggl(-\,\vec{\Delta}^2\biggr) \, ,\\
	\label{Eq:shear-1}
    	s(r) &=& 
	\frac{1}{4}\int\frac{\di^3\Delta}{2E(2\pi)^3}
	\;e^{i\vec{\Delta}\vec{r}}\;D(t)\,
    	\biggl(-\vec{\Delta}^2+3\,(\vec{e}_r\vec{\Delta})^2\biggr)\,.
}
If we choose the coordinates in the $\Delta$-integration such that 
$\vec{r}$ points along the direction of the $\Delta_z$-axis and
define $\vec{e}_r\vec{\Delta}=\cos\theta_\Delta|\vec{\Delta}|$
then, recalling that $t=-\vec{\Delta}{ }^2$ in Breit frame,
\subba{
	\label{App-Eq:pressure-2}
    	p(r) &=& 
	\frac{1}{3}\int\frac{\di^3\Delta}{2E(2\pi)^3}
	\;e^{i\vec{\Delta}\vec{r}}\;P_0(\cos\theta_\Delta)\biggl(t\,D(t)\biggr) \,,\\
	\label{App-Eq:shear-2}
    	s(r) &=& 
	\frac{3}{4}\int\frac{\di^3\Delta}{2E(2\pi)^3}
	\;e^{i\vec{\Delta}\vec{r}}\;P_2(\cos\theta_\Delta)\biggl(t\,D(t)\biggr) \,,
}
The expressions (\ref{App-Eq:pressure-2},~\ref{App-Eq:shear-2}) can be further
simplified. Using the expansion of a plane wave in spherical Bessel functions
and the orthogonality relation of Legendre polynomials,
\be
	e^{i\vec{\Delta}\vec{r}} = \sum\limits_{l=0}^\infty i^l(2l+1)\,
	j_l(|\vec{\Delta}|r)\,P_l(\cos\theta_\Delta) 
	\, , \;\;\;
	\int\limits_{-1}^1\di x\;P_l(x)P_k(x) = \frac{2}{2l+1}\;\delta_{lk}\,,
\ee
yields
\subba{
	\label{App-Eq:pressure-3}
    	p(r) &=& \phantom{-\;}
	\frac{1}{3}\int\frac{\di^3\Delta}{2E(2\pi)^3}
	\;j_0(|\vec{\Delta}|r) \biggl(t\,D(t)\biggr) \,,\\
	\label{App-Eq:shear-3}
    	s(r) &=& -\;
	\frac{1}{2}\int\frac{\di^3\Delta}{2E(2\pi)^3}
	\;j_2(|\vec{\Delta}|r) \biggl(t\,D(t)\biggr) \,.
}

It is instructive to verify the consistency of these definitions.
As a first consistency check we integrate the expression for the
energy density in Eq.~(\ref{Eq:static-EMT-T00}) over the volume
\begin{align}
 	\int\di^3r\;T_{00}(r) 
    &=	m^2\int\di^3r\int\frac{\di^3\Delta}{E(2\pi)^3}
    	\;e^{i\vec{\Delta}\vec{r}}\;
	\biggl[A(t)-\frac{t}{4m^2}(A(t)+D(t))\biggr]\nonumber\\
    &=	m^2\int\frac{\di^3\Delta}{E(2\pi)^3}
    	\;\biggl[A(t)-\frac{t}{4m^2}(A(t)+D(t))\biggr]
    	(2\pi)^3\delta^{(3)}(\vec{\Delta}\,) \nonumber\\
    &= \lim\limits_{t\to0}\,\frac{m^2}{E}
	\biggl[A(t)-\frac{t}{4m^2}(A(t)+D(t))\biggr] =  m 
\end{align}
where in the last step we used that $E = m$ for $t=-\vec{\Delta}^{\,2}\to0$, 
which yields the desired result.
As a second consistency check we integrate the pressure, as defined
in Eq.~(\ref{Eq:pressure-1}), over the volume. We obtain 
\begin{align}
    \int\di^3r\;p(r) 
    &=	\frac{1}{3}\int\di^3r\int\frac{\di^3\Delta}{2E(2\pi)^3}
    	\;e^{i\vec{\Delta}\vec{r}}\;
	\biggl[t\,D(t)\biggr]\nonumber\\
    &=	\frac{1}{3}\int\frac{\di^3\Delta}{2E(2\pi)^3}
    	\;\biggl[t\,D(t)\biggr]
    	(2\pi)^3\delta^{(3)}(\Delta)\nonumber\\
    &=	\frac{1}{3}\;\lim\limits_{t\to 0}
    	\biggl[\frac{1}{2E}\,t\,D(t)\biggr] =  0 
\end{align}
which reproduces the von Laue condition (\ref{Eq:stability}).
As a third consistency test we verify the differential equation 
(\ref{Eq:diff-eq-s-p}) connecting the pressure and shear forces.
Inserting the expressions (\ref{App-Eq:pressure-3},~\ref{App-Eq:shear-3})
into Eq.~(\ref{Eq:diff-eq-s-p}), defining $z=|\vec{\Delta}|r$, 
recalling that $t=-\vec{\Delta}{ }^2$, and using primes to denote 
derivatives of a function with respect to its argument, we obtain
\begin{align}
    &   	\frac23\;\frac{\partial s(r)}{\partial r\;}+
    		\frac{2s(r)}{r} + \frac{\partial p(r)}{\partial r\;}    
     =		\int\frac{\di^3\Delta}{2E(2\pi)^3}
		\Biggl\{
	 	\frac23\,\biggl(-\frac12\,j_2^{\,\prime}(z)\biggr) 
	+  	\frac2z\,\biggl(-\frac12\,j_2(z)\biggr)
	+  	\biggl( \frac13\,j_0^\prime(z)\biggr)\Biggr\}
	   	\,|\vec{\Delta}|\biggl[t\,D(t)\biggr] 
	= 	0 \label{App-Eq:diff-eq-s-p-CHECK}
\end{align}
which vanishes because the expression in the curly brackets is zero 
due to the identity $j_0^{\,\prime}(z) - j_2^{\,\prime}(z) - 3j_2(z)/z = 0$.

\subsection{Densities of a point-like particle}
\label{Sec-3b:densities-point-like-particle}

Let us compute the static EMT densities of a point-like Klein-Gordon
particle. With the results from Sec.~\ref{Sec-2b:free-case} we obtain 
for the energy density, pressure, and shear forces as defined in 
Eqs.~(\ref{Eq:static-EMT-T00},~\ref{Eq:pressure-1},~\ref{Eq:shear-1}) 
the results
\ba
    	T_{00}(\vec{r}\,) 
	= m^2\int\frac{\di^3\Delta}{E(2\pi)^3}\;e^{i\vec{\Delta}\vec{r}}\;
    	= \frac{m^2}{\sqrt{m^2-\vec{\nabla}{ }^2/4}\;}\;\delta^{(3)}(\vec{r}\,)
	  \,,\nonumber\\
   	p(r) 
	= \frac{1}{3}\int\frac{\di^3\Delta}{2E(2\pi)^3}
	  \;\vec{\Delta}{ }^2\;e^{i\vec{\Delta}\vec{r}}
	= -\;\frac16\;\frac{\vec{\nabla}{ }^2}
	  {\sqrt{m^2-\vec{\nabla}{ }^2/4}\;}\;\delta^{(3)}(\vec{r}\,) \,,
	  \nonumber\\
	s(r) 
	= -\,\frac{3}{4}\int\frac{\di^3\Delta}{2E(2\pi)^3}
	  \;e^{i\vec{\Delta}\vec{r}}\;
    	  \biggl((\vec{e}_r\vec{\Delta})^2-\frac13\vec{\Delta}^2\biggr)\,
	= \frac18\;\frac{\;3\,e_r^ie_r^j\,\nabla^i\nabla^j-\vec{\nabla}^2\;}
	  {\sqrt{m^2-\vec{\nabla}{ }^2/4}}\;\delta^{(3)}(\vec{r}\,) \, .
	\label{Eq:densities-with-mass-corrections}
\ea
As expected, the EMT densities of a point like particle are given 
by singular $\delta$-distributions or their derivatives. Notice that in 
Eq.~(\ref{Eq:densities-with-mass-corrections}) it is understood that 
the derivatives act only on the $\delta$-functions.

The infinite tower of derivatives implicit in the square roots is a 
consequence of what is sometimes referred to as ``relativistic corrections.''
Let us first show that despite these corrections the expressions are 
theoretically consitent. For that we assume that the square roots in
Eq.~(\ref{Eq:densities-with-mass-corrections}) can be formally expanded 
in terms of a series in powers of $\vec{\nabla}{ }^2/(4m^2)$. 
The derivatives on the $\delta$-functions are handled using 
$\int\di^3r \,h(\vec{r}\,) \nabla^i\nabla^j\delta^{(3)}(\vec{r}\,)
= [\nabla^i\nabla^jh(\vec{r}\,)]_{\vec{r}=0}$ where $h(\vec{r}\,)$
denotes a generic trial~function.
In the case of the mass $m=\int\di^3r\,T_{00}(r)$ and the von Laue
condition $\int\di^3r\,p(r)=0$ the trial functions are $h(\vec{r}\,)=1$,
and we immediately see that $T_{00}(r)$ and $p(r)$ in
Eq.~(\ref{Eq:densities-with-mass-corrections}) comply with these 
constraints. In order to verify that the $D$-term as defined in 
Eqs.~(\ref{Eq:D-from-s(r)-and-p(r)}a,~\ref{Eq:D-from-s(r)-and-p(r)}b) 
is correctly reproduced, we note that in this case the trial function 
is $h(\vec{r}\,)=r^2$ and $\nabla^i\nabla^jr^ir^j=12$ and 
$\vec{\nabla}{ }^2r^2 = 6$ holds. 
This confirms the correct result $D=-1$.

While the expressions are consistent in the above sense, the presence 
of relativistic corrections artificially mimics an internal structure. 
This can be seen, for instance, by computing the moments of the energy 
density, which we define and normalize such that the zeroth moment is 
unity (it would be the mass of the particle, had we not normalized it), 
the first moment is the mean square radius of $T_{00}(\vec{r}\,)$, etc. 
With this definition, and assuming that the expansion of the square root 
under the integral is allowed, we obtain for the moments of the energy 
density
\ba
	M_k 
	\equiv \frac1m\int\di^3r\;r^{2k}T_{00}(\vec{r}\,) 
	= \int\di^3r\;r^{2k}\;\biggl[
	\frac{1}{\sqrt{1-\vec{\nabla}{ }^2/(4\,m^2)}\;}\;\delta^{(3)}(\vec{r}\,)
	\biggr]
	= \int\di^3r\;r^{2k}\;\biggl[\sum\limits_{j=0}^\infty c_j\;
	(\vec{\nabla}{ }^{2})^j\;\delta^{(3)}(\vec{r}\,)
	\biggr] \, ,\label{Eq:T00-moments-with-rel-corr}
\ea
with $c_j = (2j-1)!!/[(4\,m^2)^j\,2^j\,j!]$ where $(-1)!! = 1!! = 1$ and 
$(2j+1)!!=1\cdot 3 \cdot \ldots \cdot (2j-1)\cdot(2j+1)$ for $j > 1$.
Performing $2j$ partial integrations in each term of the sum 
over $j$ and using 
$[(\vec{\nabla}{ }^{2})^j\,r^{2k}]_{r=0} = (2k+1)!\;\delta_{jk}$ yields
\ba
	M_k &=& \frac{(2k+1)!!\;(2k-1)!!}{(4\,m^2)^k} \; .
\ea
Let's recall that for a point-like particle one naturally expects
$M_k=\delta_{k0}$ and that $M_k\neq 0$ for $k>0$ would imply an extended
structure. This is a consequence of relativistic corrections, and a
general limitation of the interpretation of 3D-Fourier transforms of 
form factors as 3D-densities. One could also define moments of $s(r)$ 
and $p(r)$ analog to (\ref{Eq:T00-moments-with-rel-corr}) to show that
relativistic corrections do not spoil the lowest moments related to von 
Laue condition and the $D$-term, as shown in 
Sec.~\ref{Sec-3b:densities-point-like-particle}.
However, higher moments of $s(r)$ and $p(r)$ would be altered similarly 
to those of the energy density and lead to unphysical results.

The presence of relativistic corrections is of course well known, 
and their appearance can be understood in various ways, see e.g.\ 
\cite{Belitsky:2005qn} for a review.
In the next section we will discuss how (and when) one can, at least in 
principle, go about these relativistic corrections. It is important to 
notice that the relativistic corrections set limitations for the 
interpretation. Nevertheless formally all theoretical results
remain correct and consistent as we have shown above.

\subsection{``Switching off'' relativistic corrections}
\label{Sec-3c:switching-off}

In order to ``switch off'' such relativistic corrections and recover
well-defined 3D-densities consistent with the notion of a point-like
particle, let us assume from now on that we work in the heavy mass 
limit $m\to\infty$, and retain only the respectively leading terms. 
Such a description in principle applies to the (free) Higgs boson,
which is the only presently known fundamental spin zero particle.
In this way we obtain for a heavy boson
\ba
    	T_{00}(\vec{r}\,)= m
			\;\delta^{(3)}(\vec{r}\,)\,,\nonumber\\
   	p(r)=-\;\frac{\vec{\nabla}{ }^2}{6\,m}
	  		\;\delta^{(3)}(\vec{r}\,) \,, \nonumber\\
	s(r)= \frac{\;3 \;e_r^ie_r^j\,\nabla^i\nabla^j
			-\vec{\nabla}^2\;}{8\,m} \;\delta^{(3)}(\vec{r}\,) \,.
	\label{Eq:EMT-densities-large-m-limit}
\ea
One sees immediately that the expressions in 
(\ref{Eq:EMT-densities-large-m-limit}) are consistent. The von Laue condition 
(\ref{Eq:stability}) is satisfied, one obtains the same result $D=-1$ 
for the $D$-term using its both representations in terms of $s(r)$ and $p(r)$ 
in Eqs.~(\ref{Eq:D-from-s(r)-and-p(r)}a,~\ref{Eq:D-from-s(r)-and-p(r)}b), and 
the moments of the energy distribution defined in 
Eq.~(\ref{Eq:T00-moments-with-rel-corr}) satisfy $M_k=\delta_{k0}$ 
as expected for a point like particle.

An important question is:
the mass $m$ of our boson is large, but with respect to what?
This question is ill-posed in a free theory where the only dimensionfull
parameter is $m$, and the only available length scale is the Compton
wave-length of the particle $\lambda_C = 1/m$. To give a
meaning to heavy mass limit we must ``give some internal structure''
to our heavy boson. 
To take into consideration the effects of an internal structure, we 
proceed here heuristically\footnote{\label{Footnote:micro}%
	We postpone here the question how to describe such an
	``internal structure'' in terms of a microscopic dynamical 
	Lagrangian theory. This question will be addressed later
	in Sec.~\ref{Sec-4:Q-ball-log}.}
and replace the $\delta$-functions in the expressions
(\ref{Eq:EMT-densities-large-m-limit}) with suitably smeared-out 
regular and normalized functions $f(r)$,
\be\label{Eq:smearing}
	\delta^{(3)}(\vec{r}\,) 
	\to f(r) \, , \; \; \;
	I_0 \equiv \int\di^3r\,f(r)=1\,,
\ee
where it is understood that $f(r)$ reduces to a $\delta$-function
in some well-defined limit.

Let us investigate the effect of such an internal structure on
the energy density. We choose, at this point merely for illustrative 
purposes, the following representation $f_R(r)$ for the $\delta$-function
\be\label{Eq:Gaussian-representation}
	f_R(r) = \frac{1}{\pi^{3/2}R^3}\,\exp\biggl(-\frac{r^2}{R^2}\biggr)
\ee
from which we recover $f_R(r)\to \delta^{(3)}(\vec{r}\,)$ for $R\to 0$.
In the heavy mass limit using the densities in 
Eq.~(\ref{Eq:EMT-densities-large-m-limit}) the ``true'' first moment of 
the energy distribution $M_1$,  i.e.\ mean square radius of the energy
density, is given by
\be
	\la r_E^2\ra \equiv M_1 = \frac32\,R^2 \,.
\ee
Having a specific ``(toy) model'' for the energy density, we can
equally well evaluate the  mean square radius of $T_{00}(r)$ using 
the expression (\ref{Eq:densities-with-mass-corrections}) which includes 
relativistic corrections. The result we obtain and condition required for 
the interpretation in terms of 3-D densities to be applicable are as follows
\be\label{Eq:rE2-with-rel-corr}
	\la r_E^2\ra \equiv M_1=\frac{3\,R^2}{2}\biggl(1+\delta_{\rm rel}\biggr)
	\;,\;\;\;
	\delta_{\rm rel} \equiv \frac{1}{2m^2R^2}\ll 1 \,.
\ee
Thus relativistic corrections are negligible when $m^2 R^2 \gg 1$, i.e.\ 
when the Compton wave-length is small compared to the ``actual size'' of 
the particle $\lambda_C^2 \ll R^2$. We obtained this condition here in 
the context of the mean square radius of the energy density, but it
holds also for the other densities and can be derived from general
considerations \cite{Belitsky:2005qn}.

It is instructive to estimate the size of 
the corrections as defined in Eq.~(\ref{Eq:rE2-with-rel-corr}) for 
various particles, see Table~\ref{Table-corr}. For light mesons,
like pions, kaons or $\eta$ the concept of 3D-densities is clearly
not applicable. However, for heavier mesons containing charmed 
quarks the concept makes sense: e.g.\ for $\eta_c$ the relativistic
corrections are of the order of ${\cal O}(4\,\%)$. For nuclei the concept
can be safely applied: for instance for $^4$He, the lightest spin-0
nucleus, the corrections are merely of the order of ${\cal O}(0.05\,\%)$
and they diminish quickly for heavier nuclei. This can be understood in
the liquid drop model of the nucleus, where a nucleus with mass number $A$ 
has approximately the mass $\sim A \times 0.93\,{\rm GeV}$ and the radius 
$\sim A^{1/3}\times 1.2\,{\rm fm}$ which yields $\delta_{\rm rel}\sim 1.2\,A^{-8/3}$.
Although they are not spin-0 particles, we have included the proton, deuteron 
and $^6$Li in Table~\ref{Table-corr} for comparison. The concept of 
3D-densities is applicable in all 3 cases with a reasonable accuracy of 
the order of ${\cal O}(3\,\%)$ for proton, ${\cal O}(0.1\,\%)$ for deuteron,
and ${\cal O}(0.1\,\permil)$ for $^6$Li.

Notice that it is customary to speak about mean square charge radii 
also for particles like (charged) pions and kaons, even though the
concept of 3D-densities cannot be applied here. These ``radii'' are simply 
defined by the slopes of the electric form factors as, e.g.\ in the case
of the pion
\be
	F_\pi(t) = 1+\frac{\la r_{\pi,em}^2\ra}{6}\,t + {\cal O}(t^2), 
	\;\;\; \mbox{or} \;\;\;
	\la r_{\pi,em}^2\ra = 6 F_\pi^\prime(t)\biggl|_{t=0}\,.
\ee
Of course, one can introduce the concept of the ``spatial structure''
and ``size'' of pions and kaons (and other particles) without
relativistic corrections by working with 2D densities 
\cite{Burkardt:2000za,Miller:2009qu,Vall:2013ina,Miller:2010tz}. 
In that approach the 2D-radius of the particle is still related
to the slope of the form factor, but now as 	
$F_\pi(t) = 1+\frac14\,\la r_{\pi,em,2D}^2\ra\,t + {\cal O}(t^2)$
(in each case the numerical prefactor is $1/(2d_{\rm space})$ 
with $d_{\rm space}$ the number of space dimensions in the Fourier-transform).

But the concepts of pressures, shear forces and mechanical stability 
are inherently 3D.
No interpretation exists for the stress tensor in terms of 2D densities.
Therefore, if we wish to learn about the mechanical stability of nucleons 
and nuclei, we have to pay the prize of dealing with 3D densities and
the associated relativistic corrections. However, the relativistic
``blurring'' of the 3D densities for nucleons and nuclei, about
$3\,\%$ for proton and much less for nuclei, seems acceptably
small to carry on this program.

It is important to stress the different objectives of the 2D- vs
3D-density interpretations. The 2D-density description is exact and 
this is indispensable for a rigorous probabilistic partonic 
interpretation. The 3D-density description does not describe partonic
probability densities. It describes in our context mechanical response 
functions of a system. These are to be understood as 
{\it correlation functions}
which come with relativistic corrections. This approach is justified and 
gives valuable insights, as long as the corrections are acceptably small. 
As shown in Table~\ref{Table-corr}, this is the case in particular also
in the phenomenologically relevant cases of the nucleon and nuclei.

\begin{table}[t]
\begin{center}
\begin{tabular}{|l|c|c|c|l|}
\hline
  &&&&\\
\ particle 	& \ \ $J^\pi$ \ \ 
		& \ mass [GeV] \ 
		& \ \ size [fm] \ \ 
		& \ \ $\delta_{\rm rel}$ \\
  &&&&\\
\hline
  &&&&\\
\ pion	 	& $0^-$ 	&  0.14	& 0.67	& \  $2.2$ 		\\
\ kaon	 	& $0^-$ 	&  0.49	& 0.56	& \  $2.5\times10^{-1}$	\\
\ $\eta$-meson	& $0^-$ 	&  0.55	& 0.68	& \  $1.4\times10^{-1}$	\\
\ $\eta_c$-meson \ 
		& $0^-$ 	& 2.98	& 0.26	& \  $3.8\times10^{-2}$	\\
  &&&&\\
\ proton 	& ${\frac12}^+$	& 0.94	& 0.89	& \  $2.8\times10^{-2}$	\\
\ deuteron	& $1^+$		& 1.88  & 2.14	& \  $1.2\times10^{-3}$	\\
\ $^6$Li	& $1^+$		& 5.60	& 2.59	& \  $9.3\times10^{-5}$	\\
  &&&&\\
\ $^4$He	& $0^+$ 	& 3.73	& 1.68	& \  $5.0\times10^{-4}$	\\
\ $^{12}$C  	& $0^+$   	& 11.2	& 2.47	& \  $2.6\times10^{-5}$	\\
\ $^{20}$Ne  	& $0^+$   	& 18.6	& 3.01	& \  $6.2\times10^{-6}$	\\
\ $^{32}$S  	& $0^+$   	& 29.8	& 3.26  & \  $2.1\times10^{-6}$	\\
\ $^{56}$Fe 	& $0^+$   	& 52.1	& 3.74  & \  $5.1\times10^{-7}$	\\
\ $^{132}$Xe  	& $0^+$   	& 123	& 4.79  & \  $5.6\times10^{-8}$	\\
\ $^{208}$Pb	& $0^+$		& 194	& 5.50	& \  $1.7\times10^{-8}$	\\
\ $^{244}$Pu	& $0^+$		& 227	& 5.89	& \  $1.1\times10^{-8}$	\ \\
  &&&&\\
\hline
\end{tabular}
\end{center}
\caption{\label{Table-corr}%
	Masses, radii, and the sizes of relativistic corrections
	$\delta_{\rm rel}$ as defined in Eq.~(\ref{Eq:rE2-with-rel-corr}) 
	for various spin-0 mesons and nuclei. 
	Proton, deuteron, $^6$Li are included for comparison.
	Masses and mean charge radii of mesons and proton are from 
	\cite{Olive:2016xmw} except for the radii of $\eta$ taken from the 
	estimate \cite{Hodana:2012rc} and $\eta_c$ taken from the lattice 
	calculation \cite{Dudek:2006ej}.
	Nuclear masses are from \cite{Audi:2003zz} and 
	nuclear mean charge radii from \cite{Angeli-Marinova}.
	The smaller $\delta_{\rm rel}$ the more safely is applicable 
	the 3D-density interpretation of form factors.
}
\end{table}

\subsection{Stress tensor of an extended spin-0 particle}
\label{Sec-3e:extended-particle}

In this Section we investigate the stress tensor of a point-like
(heavy) boson which ``is given'' some ``internal structure.''
We continue proceeding heuristically, see Footnote~\ref{Footnote:micro},
and replace the $\delta$-function in the expressions for $p(r)$ and 
$s(r)$ in Eq.~(\ref{Eq:EMT-densities-large-m-limit}) with a suitable
regular normalized function $f(r)$ as given in Eq.~(\ref{Eq:smearing}).
We shall assume that $f(r)$ has the properties that 
(a) it is a radially symmetric function of $\vec{r}$,
(b) it is at least three times continuously differentiable,
(c) it satisfies $r^3f^{\prime\prime}(r)\to 0$ and $r^2f^\prime(r)\to 0$ 
    for $r\to 0$, and
(d) it vanishes at large distances faster than any power of $r$. 
These restrictions will be convenient in the following, even though
some of them could be relaxed (e.g.\ a large-$r$ behavior $\propto 1/r^5$ 
would be sufficient in all physically relevant situations \cite{Goeke:2007fp}
including the chiral limit).

From Eq.~(\ref{Eq:EMT-densities-large-m-limit}) we obtain the results
\ba
   	p(r)
	&=&-\;\frac{1}{6\,m}\;
	\biggl(f^{\prime\prime}(r)+\frac2r\,f^\prime(r)\biggr) \,, \nonumber\\
	s(r)
	&=&\phantom{-}\;\frac{1}{4\,m}\;
	\biggl(f^{\prime\prime}(r)-\frac1r\,f^\prime(r)\biggr) \,,
	\label{Eq:EMT-densities-extended-case}
\ea
where the primes denote derivatives with respect to the argument.
It is important that in Eq.~(\ref{Eq:EMT-densities-extended-case}) we use 
the same function $f(r)$ in the expressions for $s(r)$ and $p(r)$. This is
dictated by the conservation of the EMT, which imposes the differential
equation (\ref{Eq:diff-eq-s-p}). In fact, the relation (\ref{Eq:diff-eq-s-p})
holds exactly
\be
    	\frac23\;\frac{\partial s(r)}{\partial r\;}+
    	\frac{2s(r)}{r} + \frac{\partial p(r)}{\partial r\;} 
	= 
	\frac23\;\biggl(
		 \frac{f^{\prime\prime\prime}(r)}{4m}
		-\frac{f^{\prime\prime}(r)}{4mr}	
		+\frac{f^{\prime}(r)}{4mr^2}\biggr)
	+ \frac2r\;\biggl(
          	 \frac{f^{\prime\prime}(r)}{4m}
	  	-\frac{f^{\prime}(r)}{4mr}\biggr)	
	+ \biggl(
		-\frac{f^{\prime\prime\prime}(r)}{6m}
		-\frac{2\,f^{\prime\prime}(r)}{6mr}
		+\frac{2\,f^{\prime}(r)}{6mr^2}\biggr) 
	= 0
	\label{Eq:check-EMT-conservation}
\ee
for every function $f(r)$ which satisfies the properties a--c.
(Only here we need that $f(r)$ is 3 times continuously differentiable. For 
all other purposes 2 times continuously differentiable would be sufficient.)
Since Eq.~(\ref{Eq:check-EMT-conservation}) holds for the extended particle
and since it is equivalent to the conservation of the EMT, it is clear that 
all other properties related to the conservation of the EMT are also satisfied.
Let us show this explicitly. The von Laue condition is
\be
	\int\limits_0^\infty\di r\;r^2p(r)
	=\frac{1}{6\,m}	\int\limits_0^\infty\di r\;
	\biggl(r^2f^{\prime\prime}(r)+2r\,f^\prime(r)\biggr)
	=\frac{1}{6\,m}	\int\limits_0^\infty\di r\;
	\frac{\partial}{\partial r}\biggl(r^2f^{\prime}(r)\biggr)  
	= 0\,
\ee
for every function $f(r)$ which satisfies the properties a--c. This proves 
Eq.~(\ref{Eq:stability}). Finally, for the $D$-term of an extended particle
we obtain from the shear forces and pressure in Eq.~(\ref{Eq:diff-eq-s-p})
the unambiguous result
\subba{  \label{Eq:D-term-extended-p}
 	D 	\;
	 =	\;m \int\di^3 r\;r^2\, p(r)
	 =	-\;4\pi\int\limits_0^\infty\di r\;\biggl(
		 r^4\frac{f^{\prime\prime}(r)}{6}+r^3\frac{f^\prime(r)}{3}\biggr)
	&=&      -\;\biggl(\frac{4\cdot 3\;I_0}{6}-\frac{3\;I_0}{3}\biggr)
	 = 	-\,1\, ,\\
	\label{Eq:D-term-extended-s}
         =  	-\,\frac{4m}{15}\int\di^3r\;r^2\, s(r)
	 = 	-\;4\pi \int\limits_0^\infty\di r\;\biggl(
		r^4\frac{f^{\prime\prime}(r)}{15}-r^3\,\frac{f^\prime(r)}{15}\biggr)
	&=& 	-\;\biggl(\frac{4\cdot 3\;I_0}{15}+\frac{3\;I_0}{15}\biggr)
	 = 	-\,1\, ,
}
where we performed one or two partial integrations in the respective
terms to express the final results in terms of the integral $I_0$
introduced in Eq.~(\ref{Eq:smearing}). The conclusion is that the 
property $D= -1$ holds also for an extended boson, and this is 
guaranteed by the normalization of the function $f(r)$
in Eq.~(\ref{Eq:smearing}).

At this point a comment is in order. One must choose one and the same 
representation for $\delta^{(3)}(\vec{r})$ when smearing out the 
$\delta$-functions in the expressions for $p(r)$ and $s(r)$ in 
Eq.~(\ref{Eq:EMT-densities-large-m-limit}), 
because they are connected by the relations
(\ref{Eq:diff-eq-s-p},~\ref{Eq:check-EMT-conservation}).
However, there is no reason why we should use the same regular 
function $f(r)$ when smearing out $T_{00}(r)$. At this point of
our considerations, $T_{00}(r)$ is unrelated to $p(r)$ and $s(r)$.
This is of course an unphysical feature. The expressions for all
EMT densities should be derived from a Lagrangian of a dynamical 
theory. A non-trivial question is whether it is possible to construct 
a dynamical theory where a particle has the property $D=-1$ but is
extended and exhibits the EMT densities of a ``smeared-out point-like'' 
particle.

\begin{figure}[t!]
\includegraphics[width=5.6cm]{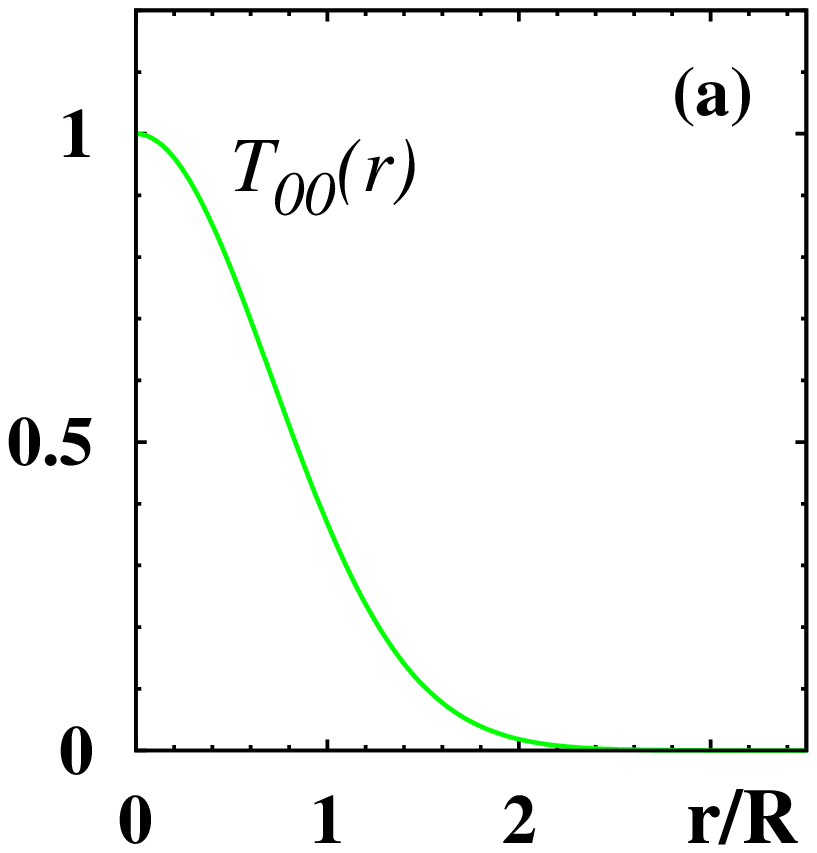}
\includegraphics[width=5.6cm]{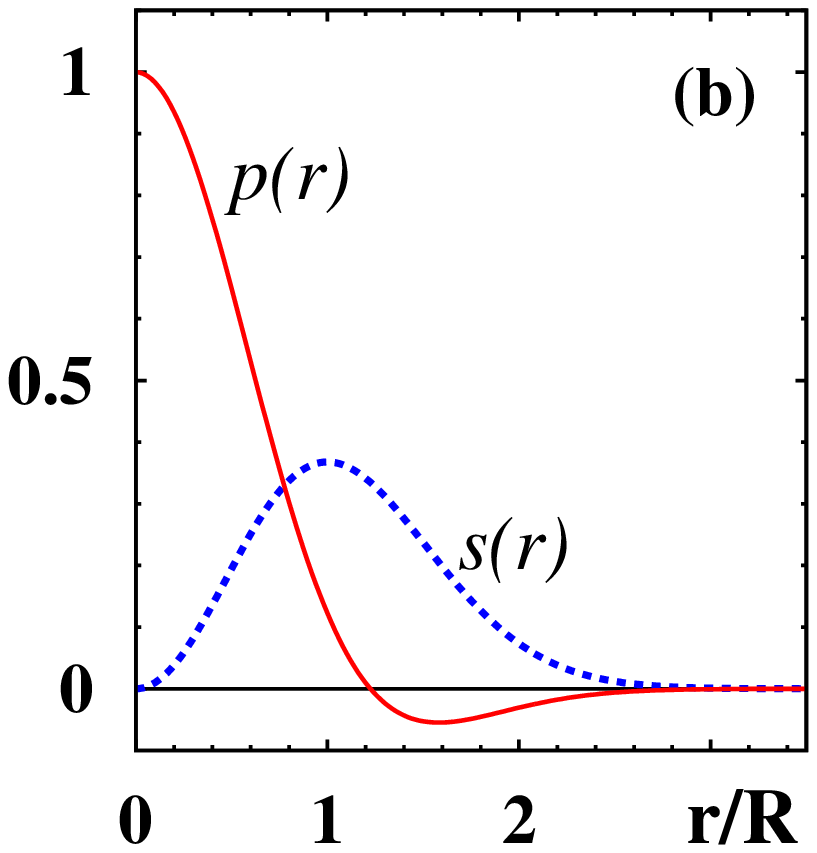} 
\includegraphics[width=5.6cm]{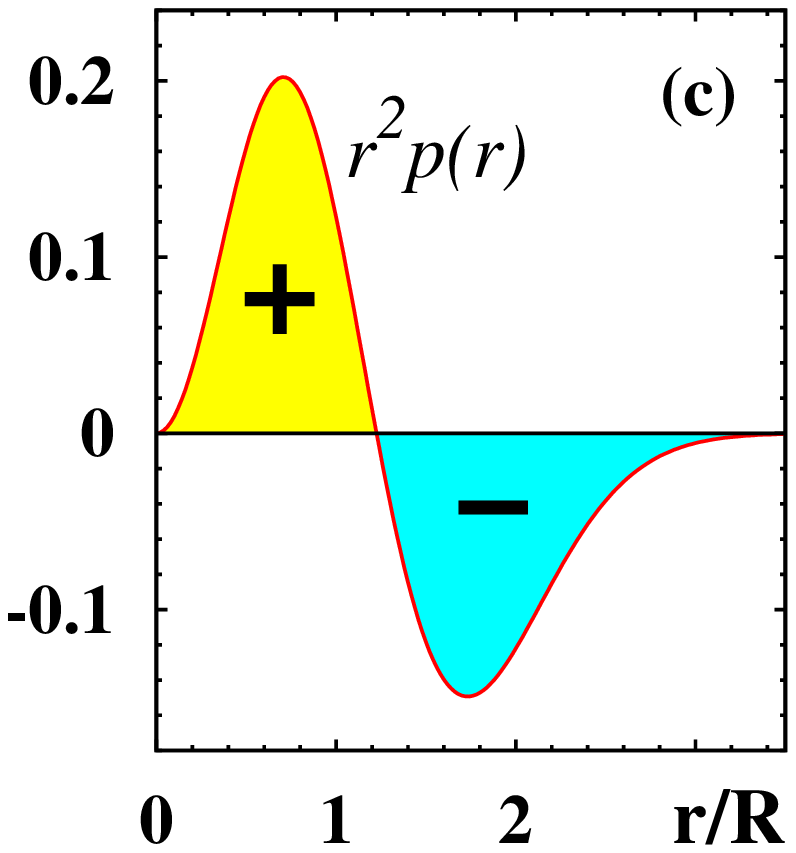}

\vspace{1cm}

\caption{\label{Fig-1:smeared-point-like-particle}
	(a) The energy density $T_{00}(r)$ in units of $T_{00}(0)$ as 
	function of $r$ in units of $R$ for a ``smeared-out'' point-particle
        from the Gaussian representation 
	(\ref{Eq:Gaussian-representation}) of a $\delta$-function.
	(b) The same as Fig.~\ref{Fig-1:smeared-point-like-particle}a
	but for $s(r)$ and $p(r)$ in units of $p(0)$. 
	(c) Visualization of the von Laue condition 
	Eq.~(\ref{Eq:stability}) with units 
	as in Fig.~\ref{Fig-1:smeared-point-like-particle}b.
	In the limit $R\to0$ (which implies $T_{00}(0)\to\infty$ and
	$p(0)\to\infty$) one recovers the original singular expressions 
	(\ref{Eq:EMT-densities-large-m-limit}). Notice that $D=-1$ holds
	both for finite $R$ as well as in the limit $R\to0$.
	For the 3D interpretation to be physically sound $R$ is required to
	be larger than the Compton wave-length of the particle.}
\end{figure}

Before addressing this question in the next section, we visualize the
EMT densities of such an ``extended particle.'' For purely illustrative
purposes, we choose the representation $f_R(r)$ for the $\delta$-function
defined in Eq.~(\ref{Eq:Gaussian-representation}). The results are shown
in Fig.~\ref{Fig-1:smeared-point-like-particle}. It is remarkable, that
in this way we effortlessly (without invoking dynamics, just by smearing 
out a point-like particle) recover the main features of the EMT densities 
calculated non-perturbatively in dynamical theories of 
$Q$-balls \cite{Mai:2012yc,Mai:2012cx,Cantara:2015sna},
chiral solitons \cite{Goeke:2007fp,Goeke:2007fq}, or
Skyrmions \cite{Cebulla:2007ei,Kim:2012ts}.

\section{Strongly interacting Gaussian scalar field}
\label{Sec-4:Q-ball-log}

The previous section has shown that the free-theory result $D=-1$ 
persisted even if the point-like spin-0 boson was given an extended 
structure. Thereby we ``introduced'' the internal structure in a 
heuristic way. The emerging question is: can one construct a 
microscopic dynamical theory in which the spin-0 particles 

(a) have an extended structure, 

(b) have the desired property $D=-1$ of a free ``point-like'' particle, and

(c) exhibit the heuristically obtained EMT densities corresponding to
``smeared $\delta$-functions'' or their derivatives?

\noindent
The answer is yes. In the following we will present one such theory, 
which can be formulated in the $Q$-ball system already mentioned in 
Sec.~\ref{Sec-2e:strong-interaction-case-Qball}. We will begin by 
briefly reviewing the description of the EMT properties of $Q$-balls 
\cite{Mai:2012yc} in Sec.~\ref{Sec-4a:Qballs-review}, and then 
show that for a specific $Q$-ball potential one deals with exactly 
our ``heuristically smeared-out point-like'' particles 
from Sec.~\ref{Sec-4b:Qball-log-potential}. 
%
%
To streamline the presentation we address technical details of this 
theory separately in Sec.~\ref{Sec-4c:boundary-cond-for-log-Qball-theory},
and discuss potential applications in Sec.~\ref{Sec-4d:Qball-log-applications}.

\subsection{Brief review of the EMT properties of $Q$-balls}
\label{Sec-4a:Qballs-review}

A brief introduction to $Q$-balls 
was already given in Sec.~\ref{Sec-2e:strong-interaction-case-Qball}.
To make this work self-contained, we review first the general $Q$-ball
properties \cite{Coleman:1985ki} including the expressions for the 
EMT densities of $Q$-balls derived in \cite{Mai:2012yc}.

The theory defined in Eq.~(\ref{Eq:Lagrangian-Qball}) of 
Sec.~\ref{Sec-2e:strong-interaction-case-Qball} admits non-topological 
solitons for a suitable potential $V$ \cite{Coleman:1985ki}. 
In their rest frame the soliton solutions are given by the expression 
quoted in Eq.~(\ref{Eq:ansatz-Qball}) with the radial field $\phi(r)$ 
obeying the equation of motion and the boundary conditions
(primes denote differentiation with respect to the argument)
\ba
&&   	\phi^{\prime\prime}(r)+\frac2r\;\phi^\prime(r)+\omega^2\phi(r)
      	- V^\prime(\phi) = 0\:,\nonumber\\
&&  	\phi(0) \equiv \phi_0 \neq 0\, ,   \;\;\;
    	\phi^\prime(0) = 0 \,,   \;\;\;
   	\phi(r)\to 0\;\;\mbox{for}\;\;r\to\infty\, .
	\label{Eq:boundary-conditions}
\ea
The global U(1) symmetry implies a Noether current with the conserved charge
\be\label{Eq:charge-Qball}
	Q = \int\di^3x\;\rho_{\rm ch}(r)\;,\;\;\;
   	\rho_{\rm ch}(r) = \omega\;\phi(r)^2\,,
\ee
whose sign is determined by $\omega$. Below we choose $\omega>0$
without loss of generality. The EMT densities read
\subba{
	T_{00}(r)&=& 	\frac12\,\omega^2\phi(r)^2+
	 		\frac12\,\phi^\prime(r)^2+V,
	\label{Eq:T00-Qball}\\
 	p(r)	&=&  	\frac12\,\omega^2\phi(r)^2-
	 		\frac16\,\phi^\prime(r)^2-V,
	\label{Eq:pressure-Qball}\\
    	s(r) 	&=&  	\phantom{\frac12\,}\phi^\prime(r)^2,
	\label{Eq:shear-Qball} 
}
The $Q$-ball densities satisfy the relation
\be\label{Eq:relation-T-p-rho-s}
     T_{00}(r)+p(r) = \omega\,\rho_{\rm ch}(r)+\frac13\,s(r)\,,
\ee
which implies the interesting $Q$-ball specific relation
\be\label{Eq:d1-from-properties}
    D = \frac49\,\biggl(
    \omega\,Q\,M\,\la r^2_Q\ra - M^2 \, \la r^2_E\ra  \biggr)\,,
\ee
with the $Q$-ball mass $M$ and mean square radii of energy 
and charge densities defined as
\ba\label{Eq:def-mean-square-radii-E-M}
	M = \int\di^3x\,T_{00}(r) \,,\;\;
     \la r^2_E\ra=\frac1M \int\di^3x\;r^2\,T_{00}(r) \,,\;\;
     \la r^2_Q\ra=\frac1Q \int\di^3x\;r^2\,\rho_{\rm ch}(r)\,.\;\;
\ea

\subsection{$Q$-balls in logarithmic potential with \boldmath $D=-1$}
\label{Sec-4b:Qball-log-potential}

To find a microscopic theory of ``smeared out'' elementary
particles, we consider $Q$-balls in the logarithmic potential
\be\label{Eq:log-potential-Qballs}
	{\cal L} = \frac12(\partial_\mu\Phi_1)(\partial^\mu\Phi_1) 
	+ \frac12(\partial_\mu\Phi_2)(\partial^\mu\Phi_2) 
	- V_{\rm log} \, , \;\;\;
	  V_{\rm log} = A\,(\Phi_1^2+\Phi_2^2)
	   - B\,(\Phi_1^2+\Phi_2^2)\,\log\biggl(C\,(\Phi_1^2+\Phi_2^2)\biggr).
\ee
This potential is not bound from below, and understood as the limiting case 
of a well-defined theory, see
Sec.~\ref{Sec-4c:boundary-cond-for-log-Qball-theory}.
Actually two parameters are sufficient to define this theory:
we can replace $C\to 1/B$ and $A\to A-B\,{\rm log}(AC)$ without
loss of generality which we shall do from now on. 
For this potential the equation of motion for the radial field reads
\be
\label{Eq:eom-Qball-III}
  	\phi^{\prime\prime}(r)
	+ \frac2r\;\phi^\prime(r)
	+ \biggl(\omega^2-2A+2B\biggr)\,\phi(r)
	+ 2B\,\phi(r)\,\log\biggl(\frac{\phi(r)^2}{B}\biggr) 
	= 0\,.
\ee
The solution satisfying the boundary conditions (\ref{Eq:boundary-conditions}) 
can be found analytically and is given by
\be	\phi(r) = \phi_0\,\exp\biggl(-B\,r^2\biggr) \,, \;\;\;
   	\phi_0  = \sqrt{B}\,\exp\biggl(\frac{2A+4B-\omega^2}{4B}\biggr)
	\,.\label{Eq:Qball-analytic-solution}
\ee
With the solution (\ref{Eq:Qball-analytic-solution}) all $Q$-ball properties
can be evaluated analytically. In particular, we obtain for the densities
\subba{
	T_{00}(r)	&=& (\omega^2-2B+4B^2r^2)\,\phi(r)^2,
	\label{Eq:T00-log-Qball}\\
 	p(r)		&=& (2B-\frac83\,B^2\,r^2)\,\phi(r)^2\,,\\
    	s(r) 		&=& 4\,B^2\,r^2\,\phi(r)^2\,,\\
   	\rho_{\rm ch}(r)	&=& \omega\;\phi(r)^2\,.\phantom{\frac11}\,
}
The expressions for $s(r)$ and $p(r)$ satisfy the general differential 
equation (\ref{Eq:diff-eq-s-p}), $p(r)$ satisfies the von Laue condition 
(\ref{Eq:stability}), and all densities comply with the $Q$-ball specific 
relation (\ref{Eq:relation-T-p-rho-s}). For the global $Q$-ball properties 
we obtain
\subba{
	Q &=& N_0\,\omega		\,,\;\;\; 
	M  =  N_0\,(B+\omega^2)		\,,\;\;\; 
	D  = -N_0^2\,(B+\omega^2)	\,,\;\;\; 
	N_0\equiv \phi_0^2\,\biggl(\frac{\pi}{2B}\biggr)^{3/2} 
	\label{Eq:global-properties-Qball-log}\\
     	\la r^2_E\ra &=& \frac{3}{4B}\,\frac{3B+\omega^2}{B+\omega^2}\,,\;\;\;
     	\la r^2_Q\ra  =  \frac{3}{4B}
	\label{Eq:radii-Qball-log}
}
It is important to stress that the same result for $D$ follows in 3 different 
ways, from Eqs.~(\ref{Eq:D-from-s(r)-and-p(r)}a,~\ref{Eq:D-from-s(r)-and-p(r)}b)
and (\ref{Eq:d1-from-properties}).
At this point it is also worth stressing that we obtain an analytic result 
for $D$ which is manifestly negative, in agreement with all available 
theoretical calculations calculations.

Next we discuss the requirements on the parameters. The conditions
(\ref{Eq:local-requirements-stability}b,
\ref{Eq:classical-stability},
\ref{Eq:local-requirements-stability}a) imply (in this order):
\subba{
	\frac23\,s(r) + p(r) = 2B \phi(r)^2 
	\ge 0 \;\;\; &\Leftrightarrow& \;\;\; B \ge 0\,,
	\label{Eq:stab-log-a}\\
	  \frac{\di}{\di\omega}	\left( 	\frac{M}{Q} \right) 
	= \frac{\di}{\di\omega}	\left( 	\omega 	+ \frac{B}{\omega}\right) 
	\ge 0	\;\;\; &\Leftrightarrow& \;\;\; \omega^2 \ge B\,,
	\label{Eq:stab-log-b}\\
	T_{00}(r) = (\omega^2-2B+4B^2r^2)\,\phi(r)^2 
	\ge 0 	\;\;\; &\Leftrightarrow& \;\;\; \omega^2 \ge 2B\,.
	\phantom{\frac11}\,
	\label{Eq:stab-log-c}
}
All conditions are satisfied and the solutions classically stable if
$2\omega^2 \ge B > 0$ (we exclude $B=0$ in (\ref{Eq:stab-log-a}) 
which would reproduce free theory).
For the limiting value $\omega^2=2B$ the energy
density vanishes in the center, which is a feature not observed so far
in the $Q$-ball literature to the best of our knowledge. 
For $2B < \omega^2 < 4B$ the energy density exhibits a dip in the center. 
Such dips occur naturally when the ``surface tension'' of the $Q$-matter 
is strong enough to produce a peak in $T_{00}(r)$ at the ``edge'' of the 
$Q$-ball \cite{Mai:2012yc}. Finally, for $\omega^2\ge 4B$ we have a 
$T_{00}(r)$ which has no dip and is monotonically decreasing for all $r$.

Notice that the parameter $A$ is completely unconstrained. We can 
choose $\sqrt{B}$ to serve as unit of mass in our theory, and 
$1/\sqrt{B}$ as length unit. Then the role of $A$ 
is to provide an overall rescaling of the fields by the factor 
$\exp(\frac12AB^{-1})$, as can be seen from (\ref{Eq:Qball-analytic-solution}). 
This implies a corresponding rescaling of the properties in 
(\ref{Eq:global-properties-Qball-log}) via $N_0\propto\exp(AB^{-1})$.
While at this point $A$ can take any value, in 
Sec.~\ref{Sec-4c:boundary-cond-for-log-Qball-theory} we 
shall will see that certain restrictions for $A$ exist.

Now we discuss how to fix the parameters such that $D=-1$.
We notice that in general for our logarithmic 
$Q$-balls 
\be
	\frac{(-D)}{Q^2} = 1 + \frac{B}{\omega^2} > 1\,,
\ee
where the inequality arises from $0<B\le2\omega^2$. Clearly, parameters can 
be chosen such that either $D=-1$ or $Q=1$ but not both simultaneously
(unless one considered a limit like $\omega\to\infty$ for fixed $B$).
However, $Q$ is a conserved but not a topological quantum number and not 
required to be an integer. It also does not need to correspond in general
to the electric charge. Notice that, if we wished to do it, we could simply 
redefine the unit in which the charges are measured to have integer-valued 
charges.
Thus, there is no principle obstacle to have $D=-1$. Notice that similarly 
$M^2=(-D)\,(\omega^2+B)$ holds, implying the nice result $M=\sqrt{\omega^2+B}$
for $D=-1$.

To obtain the desired value for the $D$-term $D=-1$ we may fix $A$ and 
$\omega$ as follows, 
\be\label{Eq:log-Qball-parameter-fixing-for-D=-1}
	\omega^2=\alpha\,B\, , \;\;\;
	A=\frac{B}{2}
	\Biggl[\alpha-4-\log\biggl[\frac{\pi^3}{8}(1+\alpha)\biggr]\Biggr]\,,
\ee
with an arbitrary positive parameter $\alpha$ which will be constrained 
shortly. In this way we obtain 
\be
	D  =  -1 					\,,\;\;\;
	M  =  \sqrt{B}\,\sqrt{1+\alpha}			\,,\;\;\; 
	Q  =  \sqrt{\frac{\alpha}{1+\alpha}} 		\,,\;\;\; 
     	\la r^2_E\ra  =  \frac{3}{4B}\,\frac{3 + \alpha}{1 + \alpha}\,,\;\;\;
     	\la r^2_Q\ra  =  \frac{3}{4B}\,.
	\label{Eq:global-properties-Qball-log-FIXED}	
\ee
For any value of $\alpha$ we have $D=-1$. Stability considerations 
(\ref{Eq:stab-log-a}--\ref{Eq:stab-log-c}) require $\alpha\ge 2$ 
leaving this parameter otherwise unconstrained.
In order to further constrain $\alpha$ we consider our criterion
(\ref{Eq:rE2-with-rel-corr}) with $R^2\to \la r_Q^2\ra$.
(We could equally well use $\la r_E^2\ra$ for that, but due to
the general relation $\la r_Q^2\ra < \la r_E^2\ra$ the criterion
is more restrictive with $\la r_Q^2\ra$.) We obtain
\be\label{Eq:delta-rel-log-Qball}
	\delta_{\rm rel} = \frac{2}{3}\,\frac{1}{1+\alpha}\,.
\ee
At this point the
parameter $\alpha$ is still not fixed, and we are free to choose 
its value to make relativistic corrections as small as we wish,
for instance choosing $\alpha>65$ guarantees $\delta_{\rm rel}<1\,\%$.

In order to close the loop and make contact with the heuristic discussion 
in Secs.~\ref{Sec-3c:switching-off} and \ref{Sec-3e:extended-particle}
we remark that the densities can be rewritten in terms of the Gaussian 
introduced in Eq.~(\ref{Eq:Gaussian-representation}) to smear out the 
$\delta$-functions as follows
\begin{subequations}
\begin{align}
	T_{00}(r)&= M\;\biggl(\frac{\alpha-2}{\alpha+1}
	+\frac{2}{1+\alpha}\,\frac{r^2}{R^2}\biggr)\,f(r)\, , \;\;\;
	\rho_{\rm ch}(r) = Q\;\,f(r)\, , \phantom{\frac11}\\
   	p(r) 	&= -\;\frac{1}{6\,M}\;
	\biggl(\frac{\partial^2}{\partial r^2}+\frac2r\,
	\frac{\partial}{\partial r}\biggr)\, f(r) \,, \;\;\;
	s(r)	 = \frac{1}{4\,M}\;
	\biggl(\frac{\partial^2}{\partial r^2}-\frac1r\,
	\frac{\partial}{\partial r}\biggr)\, f(r) \,,
	\label{Eq:EMT-densities-extended-case-II}\\
	f(r)
	&\equiv \frac{1}{\pi^{3/2}R^3}\,\exp\biggl(-\frac{r^2}{R^2}\biggr)	
	\;\;\; {\rm with} \;\;\;
	R = \frac{1}{\sqrt{2B}} \,.
\end{align}
\end{subequations}
The smeared-out $\delta$-function representation for $T_{00}(r)$ differs 
from than that of the other densities (we discussed that this 
is in general expected). Notice that $f(r)\equiv M\,\phi(r)^2$
and $\int\di^3r\,f(r)^2=1$. We can consider several limits.

In the large-$\alpha$ limit with $B$ kept fixed in 
Eq.~(\ref{Eq:log-Qball-parameter-fixing-for-D=-1}) the energy density can 
be expressed in terms of the same smeared-out function $f(r)$ which 
defines $p(r)$ and $s(r)$. In this interesting limit $D=-1$, 
$Q\to1$ and $\la r^2_i\ra \to  3/(4B)$ ($i=E,\,Q$) are fixed while the mass
grows as $M\to\sqrt{\alpha B}$ justifying the applicability of the 
3D-density description with $\delta_{\rm rel}\to0$.

Another way to implement limits is to keep $\alpha$ fixed (at a large 
enough value to keep $\delta_{\rm rel}$ in (\ref{Eq:delta-rel-log-Qball}) 
reasonably small) and take $B\to\infty$. Now we recover a heavy particle 
which becomes point-like as $\la r^2_i\ra \to 0$ in this limit. Then 
$f(r)\to\delta^{(3)}(r)$ and we literally recover the description of a 
heavy point-like particle, with $D=-1$ of course, which we wrote down 
heuristically in Eq.~(\ref{Eq:EMT-densities-large-m-limit}) 
in Sec.~\ref{Sec-3c:switching-off}. 

We consider finally the limit that $\alpha\to\infty$ and $B\to 0$ such 
that the mass $M=\sqrt{\alpha+1}\,\sqrt{B}$ remains fixed. Nothing
prevents from choosing $M$ to be moderately small or even light 
(but it must be non-zero). However, in this limit the size of our light 
particle grows since $\la r^2_i\ra \to 3\alpha/M$ which guarantees the 
smallness of $\delta_{\rm rel}$ in (\ref{Eq:delta-rel-log-Qball}) and the 
applicability of the 3D-density description. We are not aware of systems 
of this kind in particle physics, but Rydberg atoms 
(fixed and moderate mass, extremely large size) 
provide an example from atomic physics.

It is gratifying to notice that there is no way to take a limit in which one 
could recover a light and small (point-like) particle, even if one were
willing to pay the price of large relativistic corrections in 
Eq.~(\ref{Eq:delta-rel-log-Qball}). This is not surprizing: 
our very starting point was the assumption that the 3D-density description 
is applicable, so our theory does not permit to take such a limit.

The Fig.~\ref{Fig-1:smeared-point-like-particle} basically shows the 
EMT densities of our logarithmic $Q$-ball. More precisely, 
Fig.~\ref{Fig-1:smeared-point-like-particle}a shows $T_{00}(r)$ 
for $\alpha \gg 1$, while the Figs.~\ref{Fig-1:smeared-point-like-particle}b
and \ref{Fig-1:smeared-point-like-particle}c show the exact shear forces
and the pressure distributions for any $\alpha$. We recall that the results
in Fig.~\ref{Fig-1:smeared-point-like-particle} were initially obtained on 
the basis of heuristic arguments (``smearing out a point-like particle''),
and now we have derived them from a dynamical theory.

Finally, let us remark that the logarithmic potential also admits
excited states, which will be addressed elsewhere.

\subsection{Proper boundary conditions for logarithmic $Q$-ball theory}
\label{Sec-4c:boundary-cond-for-log-Qball-theory}

This section is devoted to several technical, but indispensable details 
regarding the logarithmic potential in Eq.~(\ref{Eq:log-potential-Qballs})
which is not bound from below and does not constitute an ``acceptable'' 
$Q$-ball potential in the sense of Ref.~\cite{Coleman:1985ki}.
Here we present a potential which is acceptable, bound from 
below, and contains our log-potential as limiting case.

Let us denote for simplicity $V=V(\phi)$ where $\phi=\phi(r)$ 
is the radial field. $V$ is an ``acceptable'' $Q$-ball potential if
(i) 
$V$ is two times continuously differentiable with
$V(0)=0$, $V'(0)=0$, $V''(0)=\omega_{\rm max}^2\equiv m_\Phi^2>0$, 
$V(\phi) > 0$ for $\phi\neq0$,
(ii) 
$V(\phi)/\phi^2$ has a minimum at some $\phi_{\rm min}\neq 0$
which defines the lower limit 
$\omega_{\rm min}^2=2V(\phi_{\rm min})/\phi^2_{\rm min}$
for frequencies, 
(iii) 
positive numbers $a,b,c$ exist with $c>2$ such that
$\frac12m_\Phi^2\Phi^2-V(\phi)\le {\rm min}[a,b\,|\phi|^c]$.

To construct a potential complying with the above criteria and 
containing (\ref{Eq:log-potential-Qballs}) as a limiting case,
we introduce the dimensionless parameters $0<\varepsilon_i\ll 1$ with 
$i=1,\,2$. One acceptable regular logarithmic potential $V_{\rm reg}$ 
is defined by 
\be\label{Eq:log-potential-Qballs-App-more}
	V_{\rm reg} 
	= A\,\phi^2
	+ \varepsilon_1\phi^4
	- B\,\phi^2\,\log\biggl(
	  \varepsilon_2+\frac{\phi^2}{B}\biggr) \,.
\ee
The role of the term with $\varepsilon_1\phi^4$ is to
make sure the potential is bound from below for $\varepsilon_1>0$.
The effect of $\varepsilon_2$ is to ensure a regular small field
expansion of the potential exists,  
$V_{\rm reg}=(A-B\,\log\varepsilon_2)\phi^2+{\cal O}(\phi^4)$,
which generates a finite mass term for the fundamental field.
In the limit that the $\varepsilon_i$ are negligible we recover 
the log-potential (\ref{Eq:log-potential-Qballs}).
Below we will see how this limit is understood. We begin by considering
the limiting frequencies (\ref{Eq:condition-for-existence}) 
and their difference, 
\subba{
  	\omega_{\rm max}^2 
	&=& m_\Phi^2 = [V_{\rm reg}^{\prime\prime}(\phi)]_{\phi=0} 
	 =  2A-2B\,\log\varepsilon_2  \,, \label{Eq:omega-log-max}\\
	\omega_{\rm min}^2 
	&=& \min\limits_\phi \biggl[\frac{2\,V_{\rm reg}(\phi)}{\phi^2} \biggr] 
	 =  2A+2B(1+\log\varepsilon_1-\varepsilon_1\varepsilon_2)\;,
	\label{Eq:omega-log-min}\\
	\Delta\omega^2 
	&=& \omega_{\rm max}^2 - \omega_{\rm min}^2 
	 =  2B\,f(\varepsilon_1\varepsilon_2)
	\, , \;\; 
	f(z) = z -\log z-1\,. 
	\label{Eq:omega-log-delta}
}
We first show that $\Delta\omega^2 > 0$, i.e.\ that there is 
finite $\omega$-range for solitons to exist. This is the case 
because $B>0$ holds due to (\ref{Eq:local-requirements-stability}b)
(still valid for $\varepsilon_i\ll 1$) and $f(z)>0$ for $0<z<1$.

Next we will show that $\omega_{\rm min}^2>0$ which means that 
$V_{\rm reg}(\phi)/\phi^2>0$ at its minimum. 
Notice that in the general situation the expression for $\omega_{\rm min}^2$ 
in (\ref{Eq:omega-log-min}) does not need to be positive: for given 
$A$ and $B$ one cannot have arbitrarily small $\varepsilon_1$. This
imposes a constraint on the parameters. The general condition is
\sub{
	\be
	\omega_{\rm min}^2 > 0  
	\;\;\;\Leftrightarrow\;\;\;
	\varepsilon_1\exp(1-\varepsilon_1\varepsilon_2)<\exp(-A/B)\,.
	\ee
Here we are interested in the specific situation with $D=-1$ where $A$, $B$ 
are related to each other by Eq.~(\ref{Eq:log-Qball-parameter-fixing-for-D=-1}) 
modulo negligible ${\cal O}(\varepsilon_i)$ corrections.
This implies 
	\be
	\omega_{\rm min}^2 > 0  
	\;\;\;\Leftrightarrow\;\;\;
	\varepsilon_1 > c_0
	\sqrt{\frac{\alpha+1}{e^{\alpha}}} + {\cal O}(\varepsilon_i^2)
	\, , \;\;\;
	c_0=e\,\sqrt{\frac{\pi^3}{8}}\;,
	\ee
}
i.e.\ $\varepsilon_1$ cannot be arbitrarily small. In practice, 
however, this is a loose bound as $\alpha$ must be large enough 
to ensure small relativistic corrections $\delta_{\rm rel}$,
Eq.~(\ref{Eq:delta-rel-log-Qball}). For instance, if we demand 
$\delta_{\rm rel}\lesssim 1\,\%$ then $\alpha\gtrsim66$ and 
$\varepsilon_1 \gtrsim 2.1\times 10^{-13}$. Thus $\varepsilon_1$ 
can be chosen so small that it can be neglected for practical 
purposes. Even the limit $\varepsilon_1\to 0$ can be realized for 
$\alpha\to\infty$ in which case we deal with the heavy mass limit 
of a fixed-size particle, see Sec.~\ref{Sec-4b:Qball-log-potential}. 
We remark that $\omega_{\rm min}^2 > 0$ also guarantees 
$V_{\rm reg}(\phi)>0$ for $\phi\neq0$, which ensures that $\phi=0$ is the 
correct vacuum of the theory. 

Obviously also $\omega_{\rm max}^2>0$ since
$\omega_{\rm max}^2=\omega_{\rm min}^2+\Delta\omega^2$ and we have already
proven that $\omega_{\rm min}^2$ and $\Delta\omega^2$ are both positive. 
This is also clear from (\ref{Eq:omega-log-max}) where 
(for $\varepsilon_2\ll 1$) we see that $\omega_{\rm max}^2$ is evidently 
positive and defines the mass of the $\Phi_i$-quanta.
This completes the demonstration that $V_{\rm reg}$ satisfies the
criteria (i) and (ii) of an acceptable potential.

Finally we turn to the criterion (iii), and introduce the notation
\be
	U_{\rm eff}(\phi)\equiv \frac12m_\Phi^2\phi^2-V_{\rm reg}(\phi)
	= \varepsilon_2 B^2 h(z)			\, , \;\;\; 
	h(z) = z\,\log(1+z)-\varepsilon z^2	\, , \;\;\;
	z = \frac{\phi^2}{\varepsilon_2B}	\, , \;\;\;
	\varepsilon=\varepsilon_1\varepsilon_2	\,.
\ee
The function $h(z)$ satisfies
\be\label{Eq:F}
	h(z) \le z\,\log(1+z) \le z^2 	\;\;\; \Leftrightarrow \;\;\;
	U_{\rm eff}(\phi) \le b\,|\phi|^c	\, ,\;\;\;
	b = \varepsilon_2 B^2 		\, ,\;\;\;
	c = 4\,.
\ee
This bound is useful for $\phi < \phi_{\rm eff,\,max}$ where $U_{\rm eff}(\phi)$
exhibits a maximum. For $\phi \ge \phi_{\rm eff,\,max}$ a stronger bound is
provided by $U_{\rm eff}(\phi)\le U_{\rm eff}(\phi_{\rm eff,max})$.
To determine the latter we need the extrema of $U_{\rm eff}(\phi)$ and consider 
\be\label{Eq:Fprime}
	h^\prime(z) = \log(1+z)+\frac{z}{1-z}-2\varepsilon z \stackrel{!}{=}0
\ee
which has one solution at $z=0$ corresponding to a local minimum. 
The second solution describes the global maximum at large $z\gg1$ where 
we may approximate (\ref{Eq:Fprime}) as
$h^\prime(z) = \log(z)+1-2\varepsilon z +{\cal O}(1/z^2)\stackrel{!}{=}0$ 
which is solved by
\be\label{Eq:Lambert}
	z = -\,\frac{1}{2\varepsilon}\,
	W_{-1}\biggl(-\,\frac{2\varepsilon}{e}\biggr)
	= \frac{1}{2\varepsilon}\,\log\biggl(\frac{e}{2\varepsilon}\biggr)
	+ \frac{1}{2\varepsilon}\,\log\biggl(\log
	\biggl(\frac{e}{2\varepsilon}\,\biggr)\biggr)+\dots\,\,.
\ee
$W_{-1}(x)$ denotes the inverse function of $y=x\,\exp(x)$ 
known as Lambert W-function which is defined for $x\ge -1/e$
and multivalued at negative $x$.
More precisely, $W_{-1}(x)$ denotes the branch with $W_{-1}(x)\le-1$. 
In the second step in (\ref{Eq:Lambert}) we explored the asymptotic 
expansion of $W_{-1}(x)$ for small $(-x)\to 0$ \cite{Lambert-function} 
with the dots indicating subsubleading terms.
Keeping only the leading terms we find for the position and value of
the global maximum of $U_{\rm eff}(\phi)$ the results
\be
	\phi_{\rm eff,max}^2=\frac{B}{2\varepsilon_1}\,
	\log\biggl(\frac{e}{2\varepsilon_1\varepsilon_2}\biggr)+\dots 
	\;\; , \;\;\;\;
	U_{\rm eff}(\phi_{\rm eff,max})=\frac{B^2}{4\varepsilon_1}\,
	\log^2\biggl(\frac{e}{2\varepsilon_1\varepsilon_2}\biggr)+\dots 
\ee
which shows that a maximum exists for $\varepsilon_i>0$. Thus
$U_{\rm eff}(\phi)\le {\rm min}[a,b\,|\phi|^c]$ where we can choose
$a=U_{\rm eff}(\phi_{\rm eff,max})$ and $b$, $c$ as shown in Eq.~(\ref{Eq:F}).
This completes the demonstration that also the criterion (iii)
is satisfied.

To end this section we briefly report the results of a numerical
check with the scope to investigate the size of the deviations for 
$D$ and other quantities for $\varepsilon_i\neq0$. 
We have chosen the parameters $B=2.5$, $\alpha=65$ and a common
value $\epsilon\equiv \varepsilon_1=\varepsilon_2=10^{-5}$ for sake 
of easier comparison. Recall that other $Q$-ball parameters are fixed 
by Eq.~(\ref{Eq:log-Qball-parameter-fixing-for-D=-1}) which 
ensures $D=-1$ for $\varepsilon_i\to0$. 
Let us in the following denote the additional dependence on $\epsilon$ 
of the quantities as $\phi(r,\epsilon)$,
$M(\epsilon)$, etc with $\phi(r,0)$, $M(0)$, etc corresponding to
 $\phi(r)$, $M$ in Sec.~\ref{Sec-4b:Qball-log-potential} where 
the $\varepsilon_i$ were strictly zero. To measure the deviations
we introduce $\delta\phi(r)=\phi(r,\epsilon)-\phi(r,0)$,
$\delta M=M(\epsilon)-M(0)$, etc. For the radial field we obtain
\be
	- 0.6\times10^{-3}<\frac{\delta\phi(r)}{\phi(r)} 
	< 0.3\times10^{-3}
\ee
with the largest negative deviation at small $r$ and the largest 
positive deviation around $r=$(1--2). 
For the integrated quantities we obtain
\ba
	\frac{\delta Q}{Q} = - 0.5\times10^{-3} 	\, ,\;\;\;
	\frac{\delta M}{M} = - 0.6\times10^{-3}  	\, ,\;\;\;
	\frac{\delta D}{D} =     4\times10^{-3} 	\, ,\;\;\;
	\frac{\delta \langle r_E^2\rangle}{\langle r_E^2\rangle} 
	=     3\times10^{-3}	\, ,\;\;\;
	\frac{\delta \langle r_Q^2\rangle}{\langle r_E^2\rangle} 
	=     3\times10^{-3} \,.
\ea
Let us remark that the relative accuracy of the used numerical method
is of the order $\epsilon_{\rm num}={\cal O}(10^{-7})$ which we verified 
by reproducing within such accuracy the numerical value of $D$ using 
the 3 different methods 
(\ref{Eq:D-from-s(r)-and-p(r)}a, 
\ref{Eq:D-from-s(r)-and-p(r)}b,
\ref{Eq:d1-from-properties}),
and by performing other numerical tests as described in \cite{Mai:2012yc}.

For the $D$-term we obtain for our chosen $\epsilon=10^{-5}$ the value
$D(\epsilon) = -0.995828$ instead on $-1$. Notice that we had to chose 
$\epsilon \gg \epsilon_{\rm num} ={\cal O}(10^{-7})$. Otherwise the effect 
of non-zero $\epsilon$ could not be resolved within our numerical 
accuracy. At the same time, for the chosen parameters $\alpha$, $B$
we have the theoretical constraint $\varepsilon_1 > 2.1\times 10^{-13}$
(see above). Such small $\epsilon=\varepsilon_1=\varepsilon_2$ 
can be truly neglected for all practical (numerical) purposes.
This demonstrates how our logarithmic potential 
(\ref{Eq:log-potential-Qballs}) can be practically
understood as the limiting case of the theory 
(\ref{Eq:log-potential-Qballs-App-more}).


\subsection{Potential applications in Cosmology and Beyond Standard Model}
\label{Sec-4d:Qball-log-applications}

We end this section with an exercise to get some feeling for the involved 
numbers. The only fundamental scalar particle known in the standard model 
is the Higgs boson. If we would choose e.g.\ $\alpha=99$ and our logarithmic 
$Q$-ball to have the mass of the Higgs boson, then $m_{\rm Higgs}=10\sqrt{B}$ and 
$\la r_{\rm Higgs}^2\ra^{1/2} = 0.014\,{\rm fm}$. It is not
in our scope to discuss here the phenomenology of standard model extensions 
with composed Higgs, see e.g.\ \cite{Kaplan:1983fs,Mannheim:2016lnx}.
Let us only remark that in such extensions of the standard model the Higgs 
is typically considered to be composed of new particles with masses often 
in the TeV range, implying a much smaller size 
$\sim (1\,{\rm TeV})^{-1}\sim 0.0002\,{\rm fm}$ 
compared to what our logarithmic $Q$-ball picture would suggest.
Notice, however, that this not necessarily a contradiction because the size 
dictated by the logarithmic $Q$-ball theory is not due to interactions 
with external (new physics) particles, but due to self-interactions
and the observed Higgs boson signal \cite{Olive:2016xmw} does not 
need to be incompatible with such an internal boson size.
Indeed, logarithmic potentials for a Higgs self-interaction can be 
derived naturally from beyond standard model theories \cite{Abel:2013mya} 
whereby only the Higgs self-interaction is modified, but not the couplings
to other standard model particles. The effective infra-red theory 
derived in \cite{Abel:2013mya} contains a logarithmic Higgs-mass
term analog to our effective theory (\ref{Eq:log-potential-Qballs}).
An attractive possibility is that the Higgs could be a relatively light
soliton of much heavier elementary scalar fields of a beyond-standard-model
theory.  
Finally, let us remark that logarithmic potentials have been considered
in literature, also for instance in the context of inflationary models 
driven by logarithmic potentials \cite{Barrow:1995xb}, or baryogenesis 
in minimally supersymmetric extensions of the standard model
\cite{Kusenko:1997si,Enqvist:1997si,Kasuya:1999wu}.
Such logarithmic potentials have to be understood as effective
potentials which can be generated, for instance, radiatively
\cite{Coleman:1973jx,Gildener:1976ih}.

\newpage
\section{Conclusions}
\label{Sec-4:conclusions}

We have presented a study of the EMT form factors in spin-0 systems.
Particular emphasis was put on the $D$-term, 
an interesting but so far experimentally unknown particle property
\cite{Polyakov:1999gs}, which plays the key role in accessing 
information on the internal forces inside extended particles such as 
nucleon and nuclei \cite{Polyakov:2002yz}.
Our study has focused on free, weakly and strongly interacting theories, 
and revealed that the $D$-term is the particle property which is most 
strongly dependent on the dynamics of the theory.

As a starting point we studied the $D$-term in free field theory, and
showed that the free Klein-Gordon theory makes the unambiguous prediction
$D=-1$. This result, originally obtained by Pagels in 1965 \cite{Pagels}
and largely overlooked in recent literature, is analog to the prediction 
$g=2$ for the anomalous magnetic moment from the Dirac equation.
 
We illustrated the particular sensitivity of the $D$-term to the dynamics 
by exploring the $\Phi^4$ theory. Neither the mass nor the spin are 
affected by introducing a weak $\Phi^4$ interaction in the free theory.
But the $D$-term is changed from its free theory value  $D=-1$ to $-\frac13$ 
no matter how infinitesimally weak the interaction due to
renormalization~\cite{Collins:1976vm}
(assuming the mass is renormalized such that it coincides
with its counterpart in the classical Lagrangian).

Interestingly in QCD the Goldstone bosons of spontaneous chiral symmetry 
breaking have the $D$-terms $D=-1$ in the soft pion limit, just as in 
free field theory. This is a non-trivial consequence of chiral symmetry 
breaking \cite{Novikov:1980fa,Voloshin:1980zf}. On the basis of results
from literature \cite{Donoghue:1991qv} we estimated the $D$-terms of pions, 
kaons, $\eta$-mesons which are numerically close to $D=-1$.
In general, however, in strongly interacting theories one may encounter 
sizable (always negative) values $|D|\gg 1$ for the $D$-terms, as we have 
shown by reviewing results from nuclei \cite{Guzey:2005ba} and $Q$-balls 
\cite{Mai:2012yc,Mai:2012cx,Cantara:2015sna}.

The deeper reason why the $D$-term is more strongly sensitive to 
dynamics than mass and spin is because the latter are related to 
operators of the Poincar\'e group, which imposes rigid constraints.
The $D$-term is in spin-0 (and spin-$\frac12$) systems the only 
quantity related to the EMT with no constraint due to generators 
of the Poincar\'e group. For this reason the $D$-term offers a unique 
and sensitive probe of the dynamics. 
Although the mass itself is of course also the result of dynamics, 
nevertheless the observation is that $D$ exhibits a far stronger 
sensitivity to dynamics, as is exemplified by our insights from  
``switching on'' interactions in $\Phi^4$ theory and supported
by many studies.

The second important focus of this work was the interpretation of EMT 
form factors in terms of 3D-densities giving insights on the stress 
tensor and ``mechanical forces'' inside composite particles
\cite{Polyakov:2002yz}. Again we started from the free theory, tested 
the formalism by applying it to a point-like particle, and showed the 
internal consistency of the 3D-description. This description is justified
in the heavy mass limit which requires the introduction of an additional 
scale, the size of a particle. We quantified the corrections to this 
picture and found that they are reasonably small for a particle with 
the mass and size of the nucleon, and safely negligible even for the
lightest nuclei.

We showed that the free theory result $D=-1$ persists even when 
the spin-0 boson is not point-like but given ``some internal structure.'' 
For that we heuristically ``smeared out'' the point-like particle
solution, and showed that the resulting description is consistent. 
We constructed a microscopic theory where the ``giving'' of an 
internal structure to a particle is implemented dynamically. 
This theory allows us to ``interpolate'' between extended and point-like
particle solutions with the latter emerging in a certain parametric limit.
The interaction in this microscopic theory is given by a logarithmic potential. 
Interactions of such type have been explored in literature in various 
contexts including beyond standard model phenomenology, Higgs physics 
and cosmology. Remarkably, this theory can be solved analytically. The 
solution is a non-topological soliton of $Q$-ball type \cite{Coleman:1985ki}
which, when formulated in its rest frame in terms of a complex scalar field, 
is of the type $\Phi(t,\vec{x})= \Phi_0\,\exp(i\,\omega\,t)\,\exp(-r^2/R^2)$,
i.e.\ a Gaussian.

We stress that we use the 3D-density approach as a framework to
interpret mechanical {\it response functions} of a system: the 
stress tensor, shear forces and pressure are inherently 3D concepts. 
The interpretation of such response functions in terms of 3D-densities 
remains to be taken with a grain of salt due to relativistic corrections. 
In the case of the phenomenologically interesting nucleon and nuclei such 
corrections are, however, acceptably small to allow us to carry on 
this program and gain valuable insights into internal forces.

A derivation of a 2D interpretation of the $D$-term in terms 
of lightcone densities was beyond the scope of this work. Such an
interpretation, which would be free of relativistic corrections 
\cite{Burkardt:2000za} and shed new light on the $D$-term, 
remains to be addressed in future studies.

This work contributes to a better understanding of the $D$-term, which 
has emerged already in the pre-QCD era as a fixed pole contribution 
in the angular momentum plane to the virtual Compton scattering 
amplitude in the framework of Regge theory 
\cite{Cheng:1970vg,Brodsky:1971zh,Brodsky:1972vv}
(which reflects that the $D$-term determines the 
asymptotics of GPDs in the limit of renormalization scale 
$\mu\to\infty$ \cite{Goeke:2001tz,Diehl:2003ny}, see also
\cite{Brodsky:2008qu,Muller:2015vha} for discussions).
After a first vague and inevitably model-dependent glimpse on 
the $D$-term from the HERMES experiment \cite{Ellinghaus:2002bq} 
more insights are expected \cite{JLab} on deeply virtual Compton 
scattering off nucleon \cite{Jo:2015ema} and nuclei \cite{Hattawy:2017woc}
from Jefferson Lab, COMPASS at CERN \cite{Joerg:2016hhs}, and the 
envisioned future Electron-Ion-Collider \cite{Accardi:2012qut}
which will allow us to test the theoretical understanding of
this fascinating property.

\

\noindent{\bf Acknowledgments.}
We would like to thank C\'edric Lorc\'e and Maxim Polyakov for
valuable discussions.
This work was supported in part by the National Science Foundation 
(Contract No.\ 1406298).

\

\noindent{\bf Note added.}
After this work was completed we learned of the phenomenological
study \cite{Kumano:2017lhr} were pion EMT form factors were
investigated.

\appendix

\section{Notation}
\label{App-A:notation}

There appears to be no unique notation for EMT form factors in literature.
Here are some of the used notations (on the left-hand-side of each equation)
in relation to our notation (on the right-hand-side of each equation):
\begin{alignat}{20}
&	\mbox{Ref.~\cite{Pagels}, Eq.~(8): \ \ } 		&
	\frac{G_1(q^2)}{2m^2} 	&=&    A(t), \;\;\;		&
	\frac{G_2(q^2)}{2m^2} 	&=&   -D(t), \;\;\;		&
	q^2 = t\,,\;\;\;\\
&	\mbox{Ref.~\cite{Belitsky:2005qn}, Eq.~(3.152): \ \ }	&
	\theta_2(\Delta^2) 		&=&    A(t), \;\;\;	&
	\theta_1(\Delta^2) 		&=&   -D(t), \;\;\;	&
	\Delta^2 = t\,,\;\;\;					\\
	&\mbox{Ref.~\cite{Donoghue:1991qv}, Eq.~(25): \ \ }	&
	\theta_2(q^2) 		&=&    A(t), \;\;\;		&
	\theta_1(q^2) 		&=&   -D(t), \;\;\;		&
	q^2 = t\,,\;\;\;\\
	&\mbox{Ref.~\cite{Guzey:2005ba}, Eq.~(2): \ \ }		&
	\frac{1}{2}M_A(t)	&=&    A(t), \;\;\;		&
	\frac{2}{5}d_A(t) 	&=&   -D(t). \;\;\; 
\end{alignat}
Notice also that in GPD literature, e.g.\ 
\cite{Polyakov:1999gs,Goeke:2001tz}, the notion 
of the $D$-term is used in a wider sense than in this work.
There the $D$-term is a contribution,
$D^a(z,t)$ for $a=q,\,\bar q,\,g$ with $q=u,\,d,\,\dots$ and
$z=\frac{x}{\xi}$ with support in the region $|x|\le |\xi|$,
to unpolarized GPDs. In even Mellin moments, e.g. 
$\int\di x\,x^{n-1}H^a(x,\xi,t)
	=c^a_{n,0}(t)+c^a_{n,2}(t)\xi^2+\dotso+c^a_{n,n}(t)\xi^n$
with $n$ even,
this contribution gives rise to the generalized form factors $c^a_{n,n}(t)$.
The $D^a(z,t)$ can (for the purposes of leading order evolution) 
be conveniently expanded in Gegenbauer polynomials 
with coefficients $d^a_1(t)$, $d^a_3(t)$, $\dots$ where the
$d^a_{n-1}(t)$ are related to the $c^a_{n,n}(t)$. 
In contrast to this, in our work the $D$-term is defined more 
narrowly as the form factor associated with the Lorentz structure 
$(\Delta^\mu\Delta^\nu-g^{\mu\nu}\Delta^2)$ in the Lorentz decomposition 
of the matrix elements of the total EMT operator.
Our $D(t)$ coincides with $\frac45d_1(t)=\frac45\sum_ad_1^a(t)$ 
in the notation of \cite{Goeke:2001tz}.

\section{\boldmath $D$-term of point like particle from 3D densities}
\label{App-B:point-like-particle-consistency}

It is instructive to ``rederive'' the result $D=-1$ of a free point-like
particle using the concept of 3D-densities and consistency considerations. 
Our starting are two natural assumptions:
(i) the EMT form factors of a free point-like particle are constant,
(ii) the energy density of a point-like particle must be given by 
$T_{00}(r)=m\,\delta^{(3)}(\vec{r}\,)$ if the particle is ``heavy''
or by the expression in Eq.~(\ref{Eq:densities-with-mass-corrections}) 
valid for any $m>0$.

The constraint $A(0)=1$ in (\ref{Eq:constraint-A(0)}) immediately implies 
with assumption (i) that $A(t)=1$ for all $t$. By the same argument 
$D(t)=D$ is of course also $t$-independent, but it value is apriori not 
known. To determine the value of $D$ we use assumption (ii) which implies
that the square bracket in the expression for $T_{00}(r)$ in 
Eq.~(\ref{Eq:static-EMT-T00}) must be a constant, 
\be
    	T_{00}(r) =
    	m^2\int\frac{\di^3\Delta}{E(2\pi)^3}
    	\;e^{i\vec{\Delta}\vec{r}}\;
	\underbrace{\biggl[A(t)-\frac{t}{4m^2}(A(t)+D(t))\biggr]}_{=\rm const}\,.
\ee
Clearly, we will recover the desired result if and only if $A(t)+D(t)=0$. 
As we already established that $A(t)=1$ these considerations immediately 
lead us to the conclusion that $D(t)=-1$, and in particular
\be
	D = -1
\ee 
for a point-like heavy particle. In this way, by imposing the abstract 
mathematical notion of a point-like particle, we recover $D=-1$ for a 
free point-like particle as a consistency condition of the 3D-density 
description. Notice that we have to explore here $T_{00}(r)$ for our 
purposes. Analog considerations of other EMT densities would not 
constrain $D$. 

The above arguments do not apply to the massless case discussed in 
footnote~\ref{footnote-massless-case} simply because our concepts 
require a massive particle. These arguments also do not apply to 
e.g.\ the $\Phi^4$-theory, because the bosons are not free there,
and similarly in other interacting theories. This explains why 
in general we obtain different $D$-terms in other theories. 
For Goldstone bosons of chiral symmetry breaking it is $D=-1$ 
in the soft pion limit,  but this cannot be ``explained'' in the 
above way: in this limit the Goldstone bosons are massless, and 
3D-density concepts are not applicable. The result $D=-1$ for
Goldstone bosons is a non-trivial consequence of chiral symmetry breaking 
and soft pion theorems.

\section{Canonical vs conformal EMT}
\label{App-C:canon-vs-conform}

In this work we have seen that the $D$-term depends on the used 
EMT definition. We encountered two definitions:
(i)~The {\it canonical} EMT which defined as the Noether current 
of space-time translations of the theory and symmetric in spin-0 case,
Eq.~(\ref{Eq:EMT-operator}). 
(ii) The {\it conformal} 
EMT which is given by (\ref{Eq:EMT-operator}) supplemented by the
improvement term (\ref{Eq:EMT-improved-II}) which, in the limit 
where all dimensionfull parameters in a Lagrangian are taken to zero, 
ensures conformal symmetry at classical level (which is broken in 
many theories by quantum corrections and renormalization).

In the free massless case it is necessary to work with the conformal EMT, 
because this theory is conformally invariant and the improvement term
(\ref{Eq:EMT-improved-II}) is essential to preserve this property, see 
footnote~\ref{footnote-massless-case}. The massive $\Phi^4$ theory is 
not conformally invariant, but it is appropriate to use the conformal 
EMT also here because adding the improvement term renders the EMT 
operator of that theory finite, see 
Sec.~\ref{Sec-2c:weak-interaction-case-phi^4} and  \cite{Collins:1976vm}. 
For Goldstone bosons it is forbidden to use the conformal EMT as the 
improvement term would violate chiral symmetry 
\cite{Voloshin:1982eb,Leutwyler:1989tn}.
Hence in these theories it is clear, for one reason or another,
whether the canonical or the conformal EMT has to be used. 

In other cases, it might be less clear which definition of the EMT 
should be used. For instance, in the free massive theory we argued 
that it is appropriate to use the canonical EMT due to the lack of 
a unique prescription why an improvement term should be added, see 
Sec.~\ref{Sec-2c:weak-interaction-case-phi^4}. We have seen that this 
choice receives a certain support in the shape of consistency argument 
discussed in App.~\ref{App-B:point-like-particle-consistency}.
But one does not need to be convinced by the argument of
App.~\ref{App-B:point-like-particle-consistency}, and it is
legitimate to wonder what we would obtain from a conformal EMT. In the 
massive free theory case the answer is just $D=-1/3$ instead on $-1$, cf.\ 
footnote~\ref{footnote-massless-case}. 

Also the results for the $D$-term in the $Q$-ball system, 
Refs.~\cite{Mai:2012yc,Mai:2012cx,Cantara:2015sna} and 
Sec.~\ref{Sec-4b:Qball-log-potential}, were obtained from the
canonical EMT. At this point we are not aware of an argument why a conformal 
EMT should be used for these calculations. But it is instructive to explore
it for the sake of obtaining an insight on how the EMT densities of an extended 
particle might be affected by working with one or the other EMT definition. 
When the improvement term (\ref{Eq:EMT-improved-II}) is included in the 
$Q$-ball theory, then the EMT densities are altered as follows:
\sub{
\ba
    	T_{00}(r)_{\rm conformal} &=& T_{00}(r)_{\rm canonical, \;Eq.~(\ref{Eq:T00-Qball})}	 
	+ \delta_h T_{00}(r)\\
    	p(r)_{\rm conformal} &=& \;\;\;p(r)_{\rm canonical, \;Eq.~(\ref{Eq:pressure-Qball})}
	+ \delta_h p(r)\\
    	s(r)_{\rm conformal} &=& \;\;\;\,s(r)_{\rm canonical, \;Eq.~(\ref{Eq:shear-Qball})}
	+ \delta_h s(r)
\ea}
with the additional terms given, in any $Q$-ball theory with an acceptable 
(in the sense of Sec.~\ref{Sec-4c:boundary-cond-for-log-Qball-theory})
potential, by
\sub{
\ba
     \delta_hT_{00}(r) 	&=& -h\;\frac1r(r\phi(r)^2)^{\prime\prime} \,\\
	\delta_h p(r)	&=& -h\;\biggl(	
	 		   \frac13\,(\phi(r)^2)^{\prime\prime}
			 + \frac23\;\frac1r\,(\phi(r)^2)^{\prime}
			 - \frac1r\,(r\phi(r)^2)^{\prime\prime}\biggr) \,.\\
	\delta_h s(r) 	&=& -h\;\biggl(	
	 		   (\phi(r)^2)^{\prime\prime}
			 - \frac1r\,(\phi(r)^2)^{\prime}\biggr)\,.
\ea}
We see that the conformal energy density differs from the canonical one. 
But due to
\be
	\int\di^3\delta_hT_{00}(r) 
	= -4\pi\,h\int_0^\infty\di r\;r(r\phi(r)^2)^{\prime\prime}
	= -4\pi\,h\biggl[r(r\phi(r)^2)^{\prime}-(r\phi(r)^2)\biggr]_0^\infty = 0
\ee
one obtains the {\it same} mass from the conformal and canonical EMT 
for every $Q$-ball theory which is of course expected. Similarly the 
conformal pressure differs from the canonical one, but it preserves 
the von Laue condition since
\ba
	\int_0^\infty\di r\;r^2\delta_hp(r) 
	&=& -h\int_0^\infty\di r\; r^2\biggl(	
	 		   \frac13\,(\phi(r)^2)^{\prime\prime}
			 + \frac23\;\frac1r\,(\phi(r)^2)^{\prime}
			 - \frac1r\,(r\phi(r)^2)^{\prime\prime}\biggr)\nonumber\\
	&=& -h\biggl[\frac{r^2}{3}(\phi(r)^2)^{\prime}-r(r\phi(r)^2)^\prime
	 +(r\phi(r)^2)\biggr]_0^\infty = 0\,.
\ea
Thus, independently of whether we use the conformal or canonical EMT 
(for the latter the proof was given in \cite{Mai:2012yc})
to describe the internal forces, the necessary condition for stability 
is satisfied in the same way.

The conformal expressions for $s(r)$ and $p(r)$ yield the same $D$-term 
via Eqs.~(\ref{Eq:D-from-s(r)-and-p(r)}a,~b). This can be seen by taking
the difference of the expressions for $D$ from pressure and shear forces,
which yields
\be
	m \int\di^3 r\;r^2\,\delta_h p(r) 
	+ \frac{4m}{15}\int\di^3r\;r^2\, \delta_h s(r)
        = \frac45\,m\,h\,4\pi\int_0^\infty\di r\;(r^4\phi(r)\phi^\prime(r))^\prime
	= 0 \, ,
\ee
Again this is a result valid for any $Q$-ball theory. 
However, the canonical and the conformal $D$-term differ, which
is not surprizing, see Sec.~\ref{Sec-2:FFs-of-EMT}. We obtain
\be\label{Eq-App:D-conformal}
	D_{\rm conformal} = D_{\rm canonical,\;Eqs.~(\ref{Eq:D-from-s(r)-and-p(r)}a,b)} 
	+ \delta_h D \,,
	\;\;\;
	\delta_h D  = -\frac43\,h\;4\pi\; m\int_0^\infty\di r\;r^3(\phi(r)^2)^\prime 
	> 0 \, ,
\ee
where in the last step we conclude that $\delta_h D$ is positive,
because $\phi(r)^2$ is a monotonically decreasing function of $r$,
i.e.\ $(\phi(r)^2)^\prime<0$ making the integrand in 
Eq.~(\ref{Eq-App:D-conformal}) negative. 

So far we considered a general $Q$-ball theory. It is insightful to
look at our analytically solvable logarithmic $Q$-ball theory from
Sec.~\ref{Sec-4:Q-ball-log} where all results can be obtained 
analytically. The modification of the conformal as compared to
canonical densities are particularly lucid in this theory, namely
\sub{\ba
    	T_{00}(r)_{\rm conformal} &=& T_{00}(r)_{\rm canonical, \;Eq.~(\ref{Eq:T00-Qball})}
	+ 6\,h\,  p(r)_{\rm canonical, \;Eq.~(\ref{Eq:pressure-Qball})}\,,\\
    	p(r)_{\rm conformal} &=& \;\;\;p(r)_{\rm canonical, \;Eq.~(\ref{Eq:pressure-Qball})}
	\times(1- 4\,h)\,,\\
    	s(r)_{\rm conformal} &=& \;\;\;\,s(r)_{\rm canonical, \;Eq.~(\ref{Eq:shear-Qball})}
	\times(1- 4\,h)\,.
\ea}
Thus for logarithmic $Q$-balls, the modification of the energy density
is proportional to the pressure which (conformal or not) integrates to 
zero as we have seen above. This illustrates how the modified energy 
density can still yield the same $Q$-ball mass.
The modifications of pressure and shear forces result in a simple
overall prefactor $1-4h=\frac13$ (with $h=\frac16$ in $3+1$ space-time
dimensions). This explains how in the conformal case the Laue condition 
is satisfied, and why we still get the same $D$-term from pressure 
and shear forces. The value of $D$ is, however, reduced
by the factor $1-4h=\frac13$. In particular, with the
parameters (\ref{Eq:log-Qball-parameter-fixing-for-D=-1}) which
ensure $D_{\rm canonical}=-1$ we obtain $D_{\rm conformal}=-\frac13$.

Thus, in the logarithmic $Q$-ball theory the conformal EMT yields an
equally satisfactory description of EMT densities as the canonical EMT.
We checked that this really is a general feature in the $Q$-ball system.
For instance in the $Q$-ball theory with the sextic potential explored 
in Refs.~\cite{Mai:2012yc,Mai:2012cx,Cantara:2015sna} and reviewed in 
Sec.~\ref{Sec-2e:strong-interaction-case-Qball} one obtains 
qualitatively the same picture, although the relation 
$D_{\rm conformal}\,:\,D_{\rm canonical} = 1\,:\,3$ is specific to
the logarithmic $Q$-ball theory. Thus, if it became clear that
a more consistent description of EMT densities would be provided by the
conformal (instead of canonical) EMT, one could switch to that description
without sacrifying any of the insights obtained in prior works.
As mentioned at this point we have no argument why the use of the
conformal EMT could be more appropriate than the use of the
canonical EMT. One possible situation to revise this point could
occur when considering quantum corrections to the classical $Q$-ball 
solution \cite{Graham:2001hr}, which was beyond the scope of this work.


\end{document}